\documentclass[aps,prl,twocolumn,superscriptaddress,showpacs,showkeys ]{revtex4-2}

\usepackage{graphicx}                   
\usepackage{hyperref}                   
\usepackage{amsmath,amssymb}            
\usepackage{epstopdf}
\usepackage{placeins}

\usepackage{color}
\definecolor{orange}{rgb}{1,0.5,0}
\definecolor{brown}{rgb}{0.65, 0.16, 0.16}
\definecolor{phlox}{rgb}{0.87, 0.0, 1.0}

\graphicspath{{figs/}}					

\begin{document}

	\title{The Visibility Graphs of Correlated Time Series Violate Barthelemy's Conjecture for Degree and Betweenness Centralities}
	
	\author{H. Masoomy}
	\affiliation{Department of Physics, Shahid Beheshti University, 1983969411, Tehran, Iran}
	
	\author{M. N. Najafi}
	\affiliation{Department of Physics, University of Mohaghegh Ardabili, P.O. Box 179, Ardabil, Iran}
	\email{morteza.nattagh@gmail.com}

	\begin{abstract}
		The problem of betweenness centrality remains a fundamental unsolved problem in complex networks. After a pioneering work by Barthelemy, it has been well-accepted that the maximal betweenness-degree ($b$-$k$) exponent for scale-free (SF) networks is $\eta_{\text{max}}=2$, belonging to scale-free trees (SFTs), based on which one concludes $\delta\ge\frac{\gamma+1}{2}$, where $\gamma$ and $\delta$ are the scaling exponents of the distribution functions of the degree and betweenness centrality, respectively. Here we present evidence for violation of this conjecture for SF visibility graphs (VGs). To this end, we consider the VG of three models: two-dimensional (2D) Bak-Tang-Weisenfeld (BTW) sandpile model, 1D fractional Brownian motion (FBM) and, 1D Levy walks, the two later cases are controlled by the Hurst exponent $H$ and step-index $\alpha$, respectively. Specifically, for the BTW model and FBM with $H\lesssim 0.5$, $\eta$ is greater than $2$, and also $\delta<\frac{\gamma+1}{2}$ for the BTW model, while Barthelemy's conjecture remains valid for the Levy process. We argue that this failure of Barthelemy's conjecture is due to large fluctuations in the scaling $b$-$k$ relation resulting in the violation of hyperscaling relation $\eta=\frac{\gamma-1}{\delta-1}$ and emergent anomalous behaviors for the BTW model and FBM. A super-universal behavior is found for the distribution function for a generalized degree function identical to the Barabasi-Albert network model.
	\end{abstract}

	\maketitle
	
	There are many general measures for the centrality in complex networks which have been devised to quantify the role and the importance of nodes, and to identify how much control they have over the network. Consider a network in which the \textit{agents} (which are the nodes in the network) choose shortest paths for interaction to optimize the efficiency. Then a central role is granted to a node which is visited with a higher frequency in the possible interactions, which is expressed via the \textit{betweenness centrality} (\textit{load}), defined for a node $i$ as 
	\begin{equation}
		b(i)=\sum_{m\ne i \ne n}\frac{\sigma_{m,n}(i)}{\sigma_{m,n}} 
	\end{equation}
	where $\sigma_{m,n}$ ($\sigma_{m,n}(i)$) is the total number of shortest paths from node $m$ to node $n$ (through $i$). This is something different from the \textit{degree centrality} which deals with more interactive agents having higher number of connections, and is defined as $k(i)=\sum_{j}A_{ij}$, where the adjacency matrix $A_{ij}$ is $1$ when there is an edge between nodes $i$ and $j$ and zero otherwise. There are other centralities, like \textit{eigenvector centrality} and \textit{closeness centrality}, which identify the impact of nodes. The degree and betweenness centralities apply to a wide range of systems like the social networks, biology, scientific cooperation, and transport. The huge numerical~\cite{goh2001universal, goh2002classification, yan2006efficient} and analytical~\cite{szabo2002shortest, wang2008betweenness, barthelemy2004betweenness} investigation of these centralities show their usefulness in studying complex systems. The problem of centralities in the scale-free (SF) networks is much more interesting which are classified in universality classes according to their scaling exponents. To be more precise, the spectrum of the quantities follows  power-law behaviors with scaling exponents which are exploited for identifying the universality classes. These networks show a power-law distribution for the degree, i.e. $p(k)\propto k^{-\gamma}$ (up to a cutoff value), where $\gamma$ (usually in the interval $[2,3]$) is called degree exponent. A similar power-law decay is observed for the betweenness centrality $p(b)\propto b^{-\delta}$, $\delta$ being the betweenness exponent, which is not generally independent of $\gamma$. As a well-known fact for SF networks, when the conditional probability distribution $p(b|k)$ is a narrow function of both $k$ and $b$, then $b\propto k^{\eta}$ with a hyperscaling relation~\cite{vazquez2002large}
	\begin{equation}
		\eta=\frac{\gamma-1}{\delta-1}.
		\label{Eq:HyperScaling}
	\end{equation}
	It was conjectured by Goh \textit{et. al.}~\cite{goh2002classification} that the amount of $\delta$ is robust and can be used to classify SF networks. Two universality classes were proposed based on the value of $\delta$, i.e. $\delta=2.2(1)$ (for the protein-interaction networks, the metabolic networks for eukaryotes and bacteria, and the co-authorship network), and $\delta=2.0(1)$ (for the Internet, the World Wide Web, and the metabolic networks for Archaea)~\cite{goh2002classification, goh2003goh}. Barthelemy argued that this conjecture is questionable since $\delta$ varies continuously as a function of $\gamma$ in many networks. Indeed, Barthelemy concluded that the only restrictions that the exponents have are~\cite{barthelemy2003comment, barthelemy2004betweenness}
	\begin{equation}
		(\text{C}\mathbb{I}):\ \eta_{\text{max}} = \eta_{\text{SFT}} = 2, \ \ (\text{C}\mathbb{II}):\ \delta \ge \frac{\gamma+1}{2},
		\label{Eq:Claim}
	\end{equation}
	where SFT represents scale-free trees. The first equation ($\text{C}\mathbb{I}$) states that $\eta$ is maximal for SFTs and the inequality ($\text{C}\mathbb{II}$) is based on Eq.~\ref{Eq:HyperScaling} for $\eta_{\text{max}}$ (the equality holds for SFTs). $\text{C}\mathbb{I}$ serves as an important difference between SF networks and SFTs. It is worthy to note that the violation of Eq.~\ref{Eq:HyperScaling} (caused e.g. by large $b$-$k$ fluctuations) leads to some anomalous behaviors, one of which is violating $\text{C}\mathbb{II}$. The large $b$-$k$ fluctuations have already been observed in some circumstances which is a source of other anomalies~\cite{guimera2004modeling,barrat2005effects,sienkiewicz2005statistical, barthelemy2011spatial}. Surprisingly, little attention has been paid to the domain of validity of the conjecture Eq.~\ref{Eq:Claim}, especially in the presence of anomalous high $b$-$k$ fluctuations, where the mean-field arguments do not have sufficient accuracy. Here we show that Barthemely's conjecture is highly restricted, i.e. it does not apply for some SF visibility graphs (VGs).\\
	
	VG is a tool to convert a given time series to a network and plays an essential role in determining the properties of nonlinear dynamical systems. Many statistical~\cite{lacasa2008time} and topological~\cite{masoomy2021persistent} aspects of VGs have been studied numerically and analytically, making it a standard powerful tool to study various systems like earthquakes~\cite{aguilar2013earthquake}, economics~\cite{rong2018topological}, ecology~\cite{braga2016characterization}, neuroscience~\cite{wang2016functional, zhu2014analysis}, and biology~\cite{zheng2020visibility}. An important step towards an understanding the scaling properties of VGs was taken by Lacasa, who showed that a self-similar time series converts into an SF network, emphasizing that the power-law degree distributions are related to the fractality~\cite{lacasa2008time,lacasa2009visibility}. The degree and betweenness centrality of VGs is of vital importance in the analysis of a time series, since they reflect the properties of hubs (rare events) of the time series under investigation.\\
	
	We consider a time series $\left\lbrace s(t_i)\right\rbrace_{i=1}^N $, where $s$ is called the \textit{activity} here, and $N$ is a maximal time in the analysis, which is the size of the VG at the same time. The VG denoted by $G(V,E)$ is a graph in which the times are the nodes (the set $V$) and $E$ is the edges connecting the nodes, so that the size of the VG is $N \equiv |V|$. The adjacency matrix for VGs is defined by~\cite{masoomy2021persistent}
	\begin{equation}
		A_{ij}=\left\lbrace \begin{matrix}
			1 & , & |t_i-t_j|=1 \\
			\prod_{k=i+1}^{j-1}\Theta(s_{ij}-s_{ik}) & , & |t_i-t_j|>1
		\end{matrix}\right. 
	\end{equation}
	where $s_{mn}\equiv \frac{s(t_n)-s(t_m)}{t_n-t_m}$ shows slopes, and $\Theta$ is a step function.
	\begin{figure}
		\includegraphics[width=80mm]{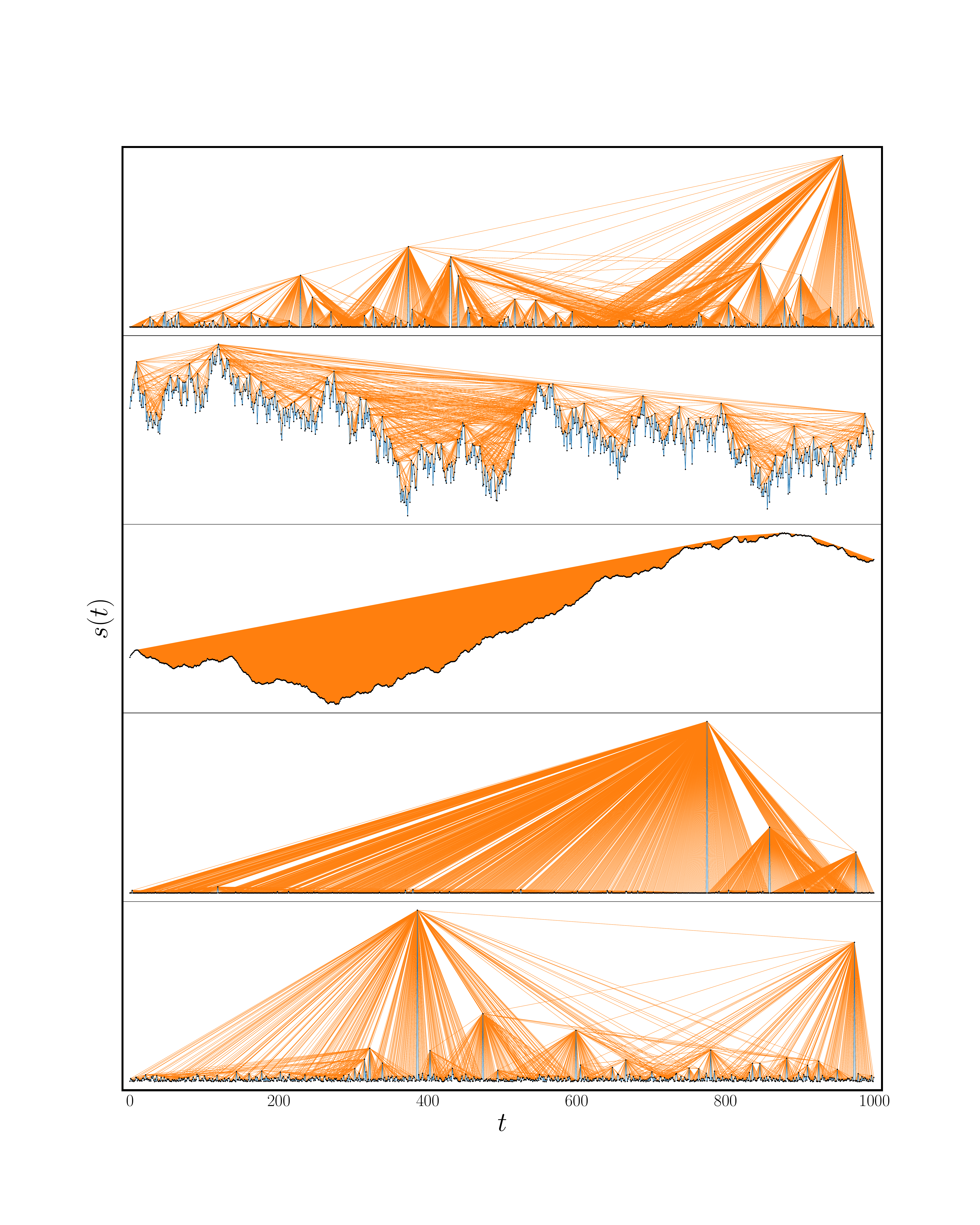}
		\caption{Visibility graphs constructed from the time series of various process studied in this work. BTW model (top panel), FBM series $H=0.2$ (middle-top panel), FBM series $H=0.8$ (middle panel), Levy walk $\alpha=0.9$ (middle-bottom panel) and Levy walk $\alpha=1.6$ (bottom panel).}
		\label{fig:nvg}
	\end{figure}
	The knowledge of the statistical properties of VGs is limited to very limited cases. It is believed that for the VG of fractional Brownian motion (FBM) and fractional Gaussian noise (FGN), $\gamma$ varies linearly by the Hurst exponent as~\cite{lacasa2009visibility} $\gamma_{FBM}(H)=3-2H$ and $\gamma_{FGN}(H)=5-2H$ for FBM and FGN respectively, which is improved further in~\cite{ni2009degree} by $\gamma_{FBM}(H)=3.35-2.87H$. Other statistical observables like the clustering coefficient, mean length of the shortest paths and motif distribution, as well as assortative mixing pattern, are studied in~\cite{xie2011horizontal}. The homological properties of weighted VG of FGN were considered in~\cite{masoomy2021persistent}, where it was shown that the persistence entropy behaves logarithmically with the size, and more importantly, the VGs are the topological tree. \\
	
	Here we systematically study the VGs for three following general processes which are representatives of leading important classes in statistical mechanics as well as the nonlinear systems:
	\begin{itemize}
		\item \textit{The BTW sandpile model,} as a  prototypical example of self-organized critical systems which show criticality without tunning of external parameters.
		\item \textit{One-dimensional (1D) FBM}, which is a popular model for both short-range dependent and long-range dependent phenomena in various fields, including physics, biology, hydrology, network research, financial mathematics etc~\cite{nourdin2013cross}, which explains why we consider this class of correlated time series. 
		\item \textit{The 1D Levy walks}, as a prototype of time-correlated self-similar systems, defined by random walks for which the step size ($s$) follows from a power-law probability density function~\cite{applebaum2009levy}
		\begin{equation}
			p(s)\propto s^{-1-\alpha}
			\label{Eq:power_law}
		\end{equation}
		where $\alpha$ is the \textit{step index} tuning the correlations. 
	\end{itemize}

	\textit{\textbf{The definition of the models:}} \\
	
	The BTW model is defined based on avalanche dynamics as the main ingredient of many natural systems, like real sandpiles~\cite{dickman2001unpublished}, earthquakes~\cite{bak1989earthquakes,rahimi2021multifractal}, sun flares~\cite{charbonneau2001avalanche}, forest fire~\cite{turcotte2004landslides}, clouds~\cite{najafi2021self,lohmann2016introduction}, Barkhausen effect in superconductors~\cite{najafi2020geometry}, rainfall~\cite{peters2001complexity}, for a good review see~\cite{najafi2021some}. In this model one initially attributes to each site $i$ of a square $L\times L$ lattice a random height $h_i\in [1,4]$, and adds a grain to a random site $i$ so that $h_i\rightarrow h_i+1$. The site $i$ is called \textit{unstable} if $h_i>4$ after which a toppling takes place according to which four grains leave the site $i$ and each neighboring site rise by one unit. We add grains one by one, so that we pass the \textit{transient configurations} and reach the \textit{recurrent configurations} identified as the state for which the average height becomes nearly constant. Each avalanche is the chain of activities between two successive stable configurations, with the size $s$ defined as the total number of local relaxations in an avalanche. For more details see~\cite{najafi2021some}. \\
	
	FBM is controlled by a Hurst exponent $H$ defined by the relation
	\begin{equation}
		\left\langle s_H(t)s_H(t')\right\rangle =\frac{1}{2}\left[|t|^{2H}+|t'|^{2H}-|t-t'|^{2H} \right], \end{equation} 
	where $s_H$ is the FBM, and $\left\langle \right\rangle $ means ensemble average. It is obtained using the standard relation $s_H(t)=\frac{1}{\Gamma(H+1/2)}\int_0^t\left(t-t' \right)^{H-1/2}\text{d}s_{H=0.5}(t')$, where $s_{H=0.5}(t)$ is the standard 1D Brownian motion. \\
	
	The Levy distribution (for which the central limit theorem does not hold in its standard form) has a long-range algebraic tail according to Eq.~\ref{Eq:power_law} corresponding to large but infrequent steps, so-called \textit{rare events}. For $\alpha>2$ the mean square deviation (MSD) of the step distribution is finite, and therefore according to the central limit theorem, the dynamic exponent locks onto $2$, corresponding to ordinary diffusive behavior. In the opposite case, however, for $\alpha<2$ MSD diverges, and the dynamic exponent equals $\alpha$ (superdiffusion), for which the dominant behavior is dictated by the rare events in long times~\cite{applebaum2009levy}. For the FBM and Levy walks we used \textit{fbm} and \textit{SciPy} Python packages, respectively. \\
	
\begin{figure*}
		\includegraphics[width=55mm]{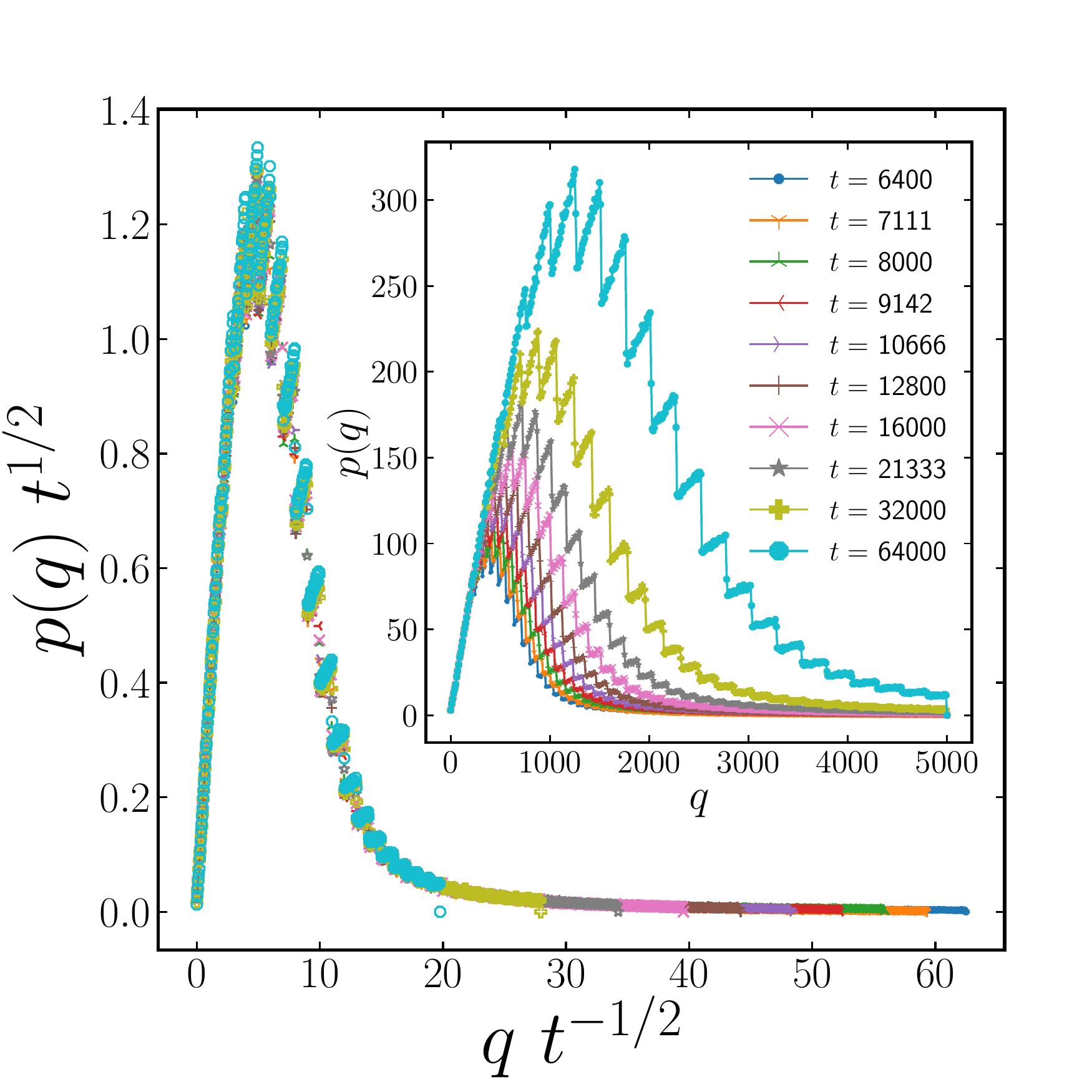}
		\includegraphics[width=55mm]{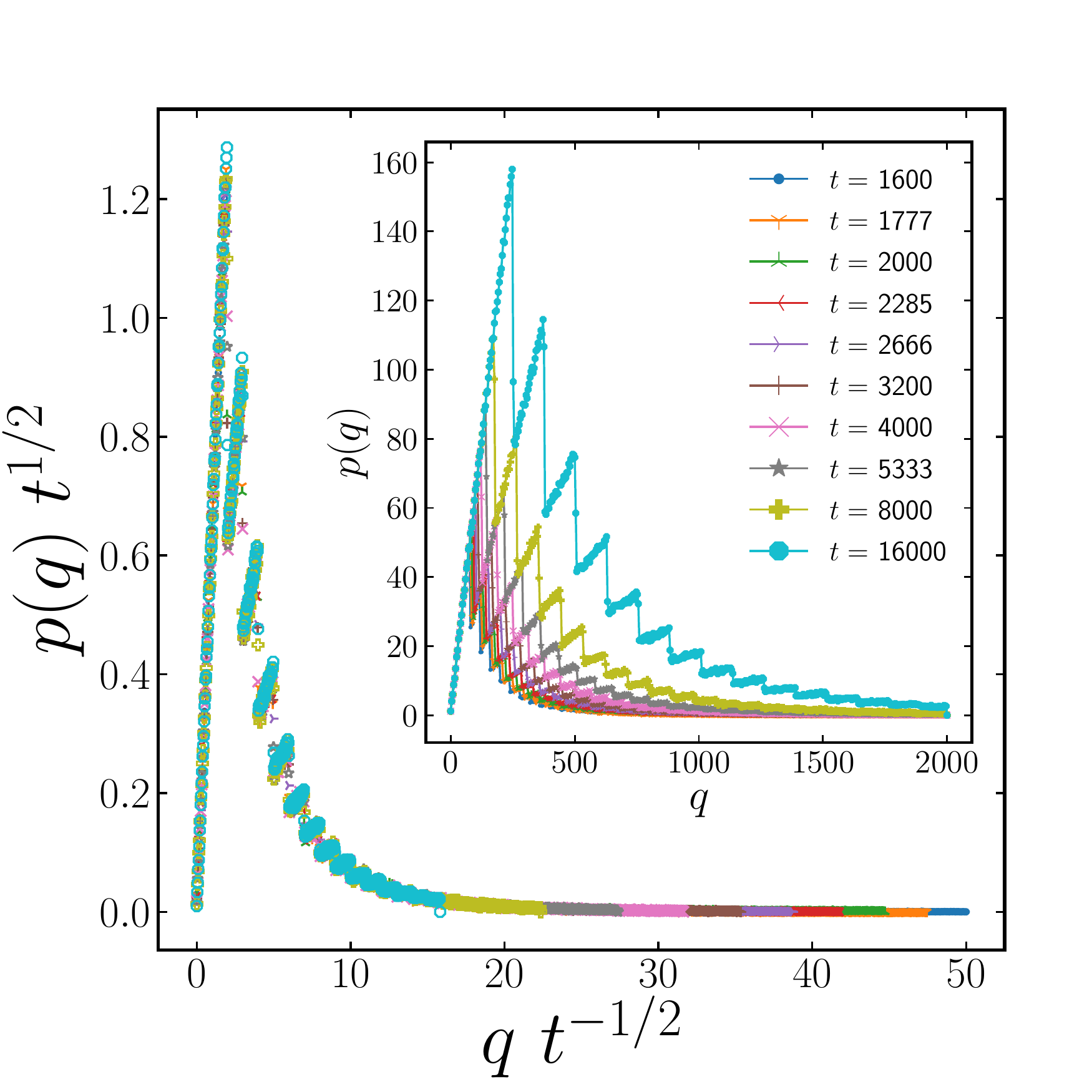}
		\includegraphics[width=55mm]{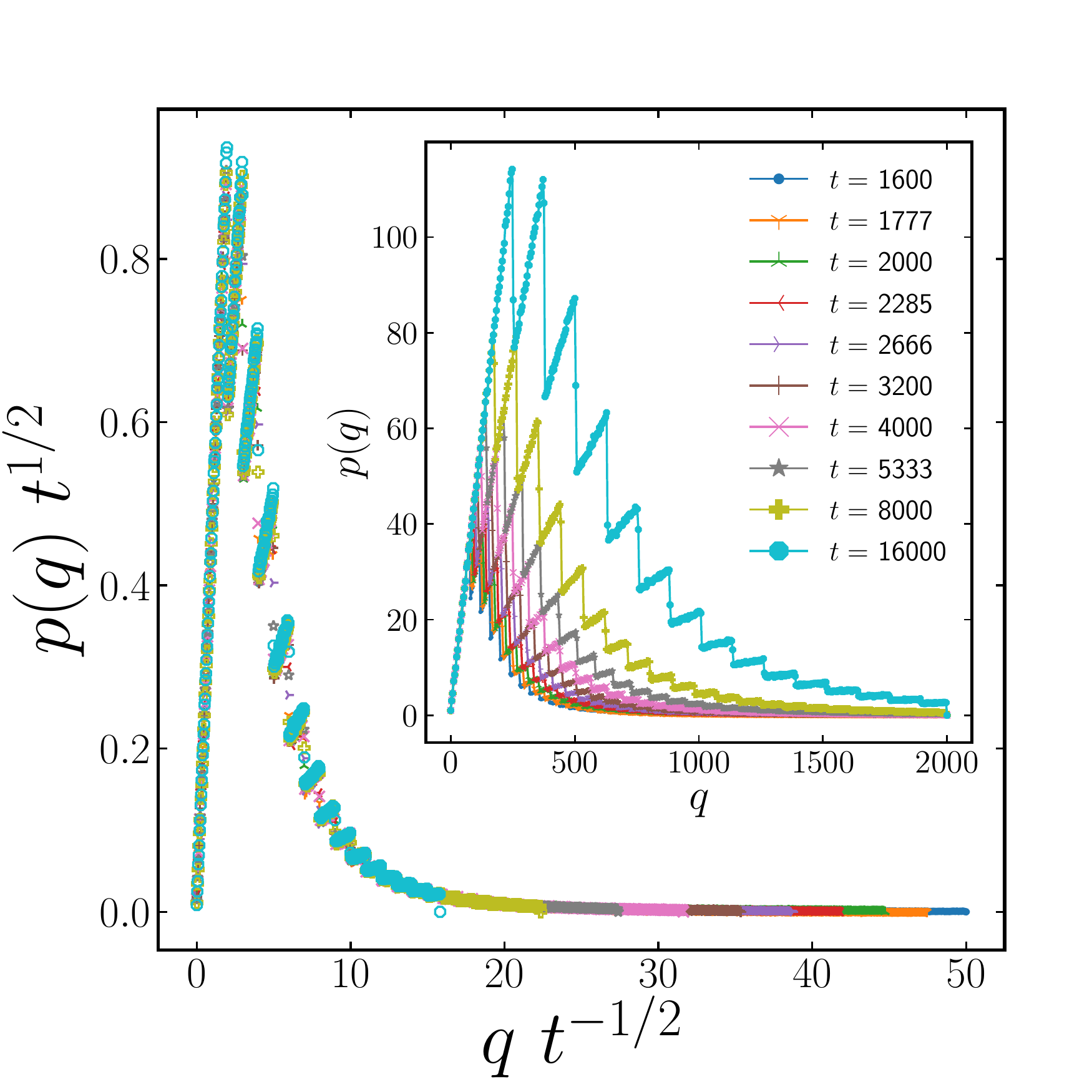}
		\includegraphics[width=55mm]{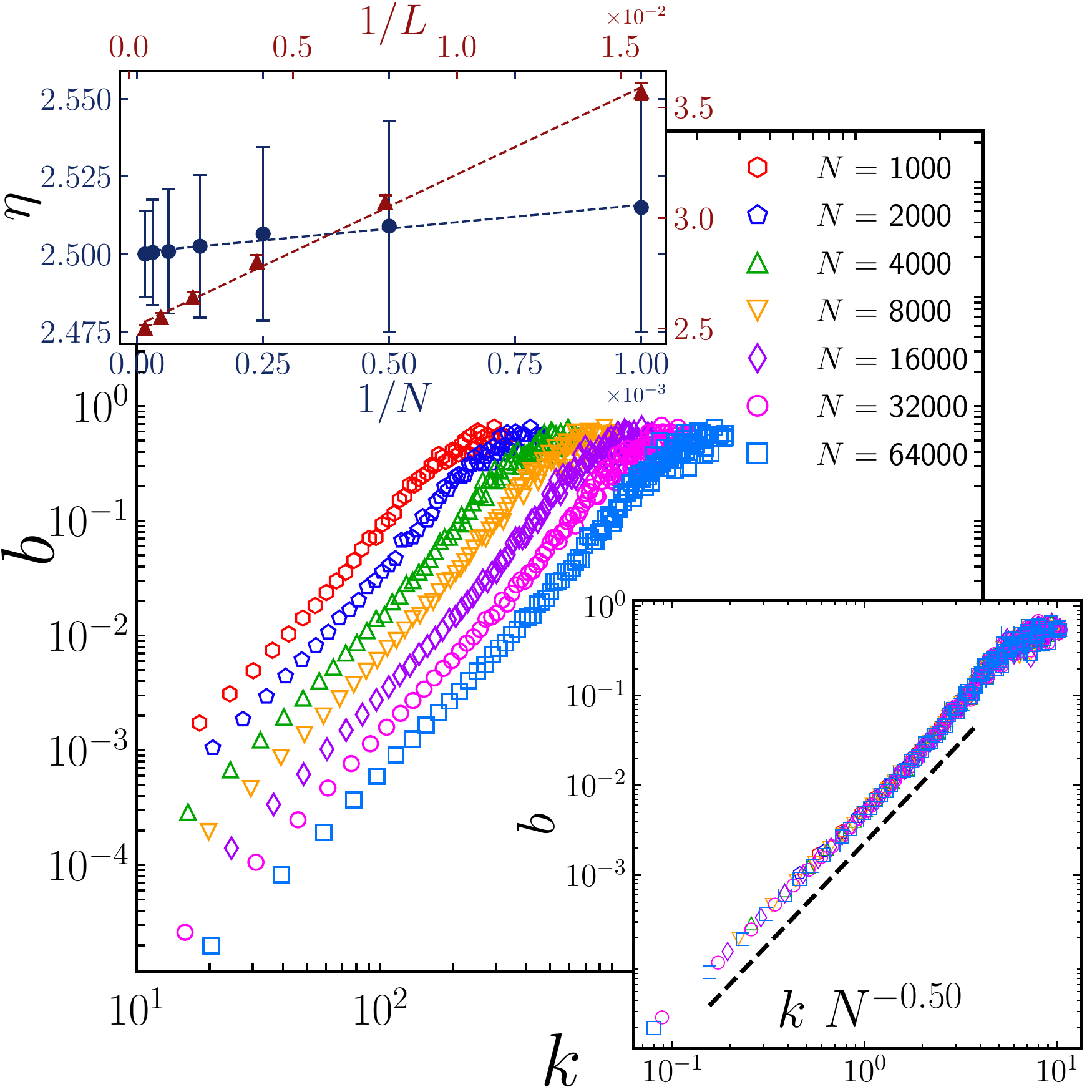}
		\includegraphics[width=55mm]{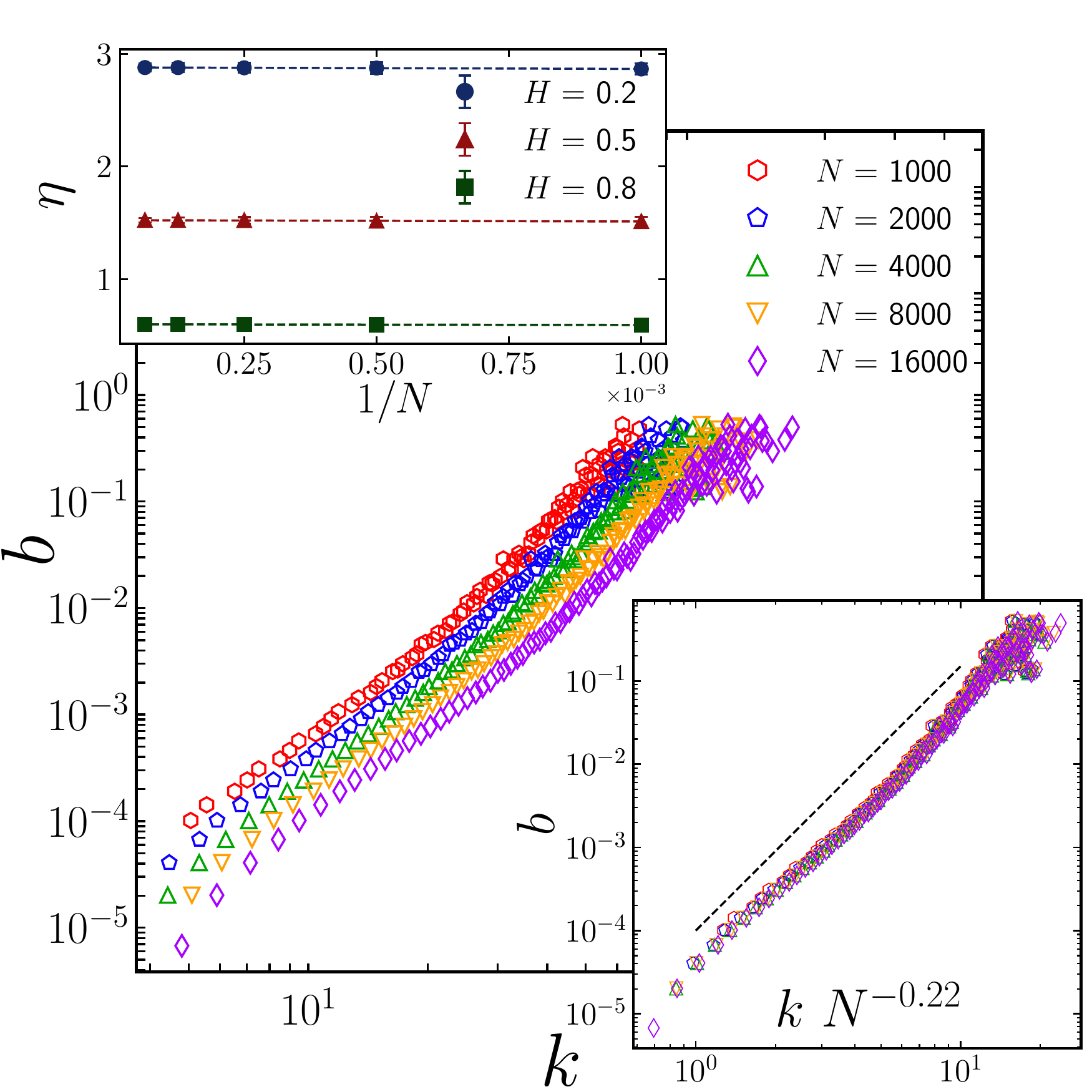}
		\includegraphics[width=55mm]{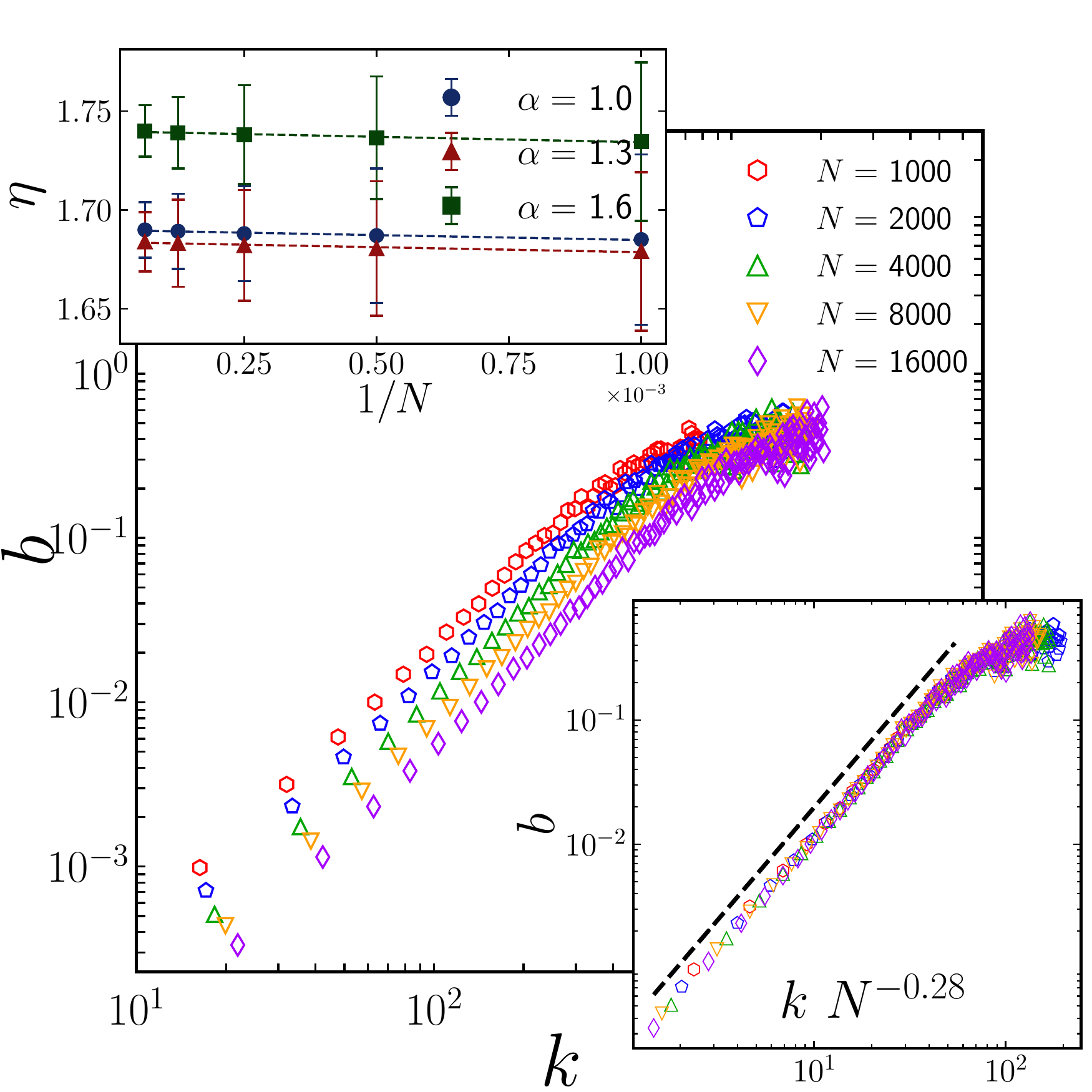}
		\caption{Top row: time-dependent probability distribution function of generalized degree for visibility graphs constructed by BTW model (left), FBM series (middle) and Levy walk (right). Bottom row: averaged betweenness centrality of nodes versus their degree in log-log scale for visibility graph constructed from BTW model (left), FBM series (middle) and Levy walk (right). The top insets indicate size dependency of the exponent $\eta$ (network size $N$ and lattice size $L$ for BTW model and $N$ for the FBM series and Levy walk), while the bottom insets are showing the data collapse of the main plots.}
		\label{fig:Graphs}
	\end{figure*}

	We simulated the BTW model for $L=64,128,256,512,1024$ and $2048$ and for all of the models we considered $\frac{N}{10^3}=1,2,4,8$ and $16$ (for the BTW model $\frac{N}{10^3}=32,64$ are added). Some VG samples are shown in Fig.~\ref{fig:nvg} for BTW, FBM$_{H=0.2}$, FBM$_{H=0.8}$, Levy$_{H=0.9}$, and Levy$_{H=1.6}$. An important check in growing SF networks is concerning their dynamic scaling properties, helping to identify their universality classes. We first consider the generalized degree function $q_i(t)\equiv \sqrt{t_{i}^{\text{birth}}}k_i(t)$, where $t_{i}^{\text{birth}}$ is the birth time of the node $i$. It is well-know that for the Barabasi-Albert (BA) network the dynamic distribution function of $q$ satisfies~\cite{hassan2011dynamic}
	\begin{equation}
		p(q,t)=t^{-\frac{1}{2}}F_{\rm BA}\left(t^{-\frac{1}{2}}q \right) 
		\label{Eq:generalizedQ}
	\end{equation}
	where $F_{\rm BA}(x)$ is a universal function with $F_{\rm BA}(x)\propto x^2$ for $x<1$ and $\propto \exp -1.4x$ for $x>2$ for $m=1$, and $F_{\rm BA}(x)\propto x^{2.9}$ for $x<1$ and $\propto \exp -2.5x$ for $x>1.5$ for $m>1$ ($m$ is the number of links that are constructed upon adding a new node, and for $m=1$ the BA network is a tree)~\cite{hassan2011dynamic}. For the models considered in this paper, although the universal functions are quite different, the same dynamic scaling exponents are observed as Eq.~\ref{Eq:generalizedQ}. The data collapse (re-scaled functions) are depicted in Fig~\ref{fig:Graphs} (top row). The universal functions $F_{\text{BTW}}(x)$, $F_{\text{FBM}}(x)$ and $F_{\text{Levy}}(x)$ are all linearly increasing functions of $x$ for small $x$'s, demonstrating that $p(q,t)\propto t^{-1}q$ for small $t^{-1}q$ values. Importantly, the exponents do not depend on $H$ and $\alpha$ for FBM and Levy, respectively, showing that these exponents are super-universal. \\
	
	To assess the Barthemly's conjecture we consider the behavior of $k$ as well as $b$. Figure~\ref{fig:Graphs} (bottom row) show the $b$-$k$ dependence, the insets of which show the behavior of $\eta$ in terms of system size ($N$ and $L$ for the BTW model, and $N$ for the others). First observe that the data in the $b$-$k$ diagrams are properly collapsed showing a finite size scaling $b\propto N^{-\beta\eta}k^{\eta}$ \textit{for all models}, introducing a new exponent $\beta$. These exponents are $\beta_{\text{BTW}}=0.50\pm0.03$, and $\beta_{\text{FBM}}$ and $\beta_{\text{Levy}}$ depend on $H$ and $\alpha$, respectively. Moreover, for the BTW model, $\lim_{L\rightarrow \infty} \eta_{\text{BTW}}=2.32\pm 0.02$ for fixed maximum $N$, and  $\lim_{N\rightarrow \infty} \eta_{\rm BTW}=2.32\pm 0.01$ for fixed maximum $L$. This is served as the first evidence of the failure of Bathelemy's conjecture ($\text{C}\mathbb{I}$)~\cite{barthelemy2004betweenness}, i.e. $\eta_{\text{BTW}}>\eta_{\text{max}}$. As expected from the standard theory of critical phenomena, the distribution functions for $k$ and $b$ also show power-law behaviors as argued above with the exponents $\gamma_{\text{BTW}}=2.60\pm 0.01$ and $\delta_{\text{BTW}}=1.71\pm 0.02$, respectively (a similar power-law decay ware observed for the clustering coefficient versus degree and betweenness centrality versus clustering coefficient with the exponents $\mu_{\text{BTW}}=0.956\pm 0.002$ and $\nu_{\text{BTW}}=2.77\pm 0.01$ respectively). \\
	
	\begin{figure}
		\includegraphics[width=80mm]{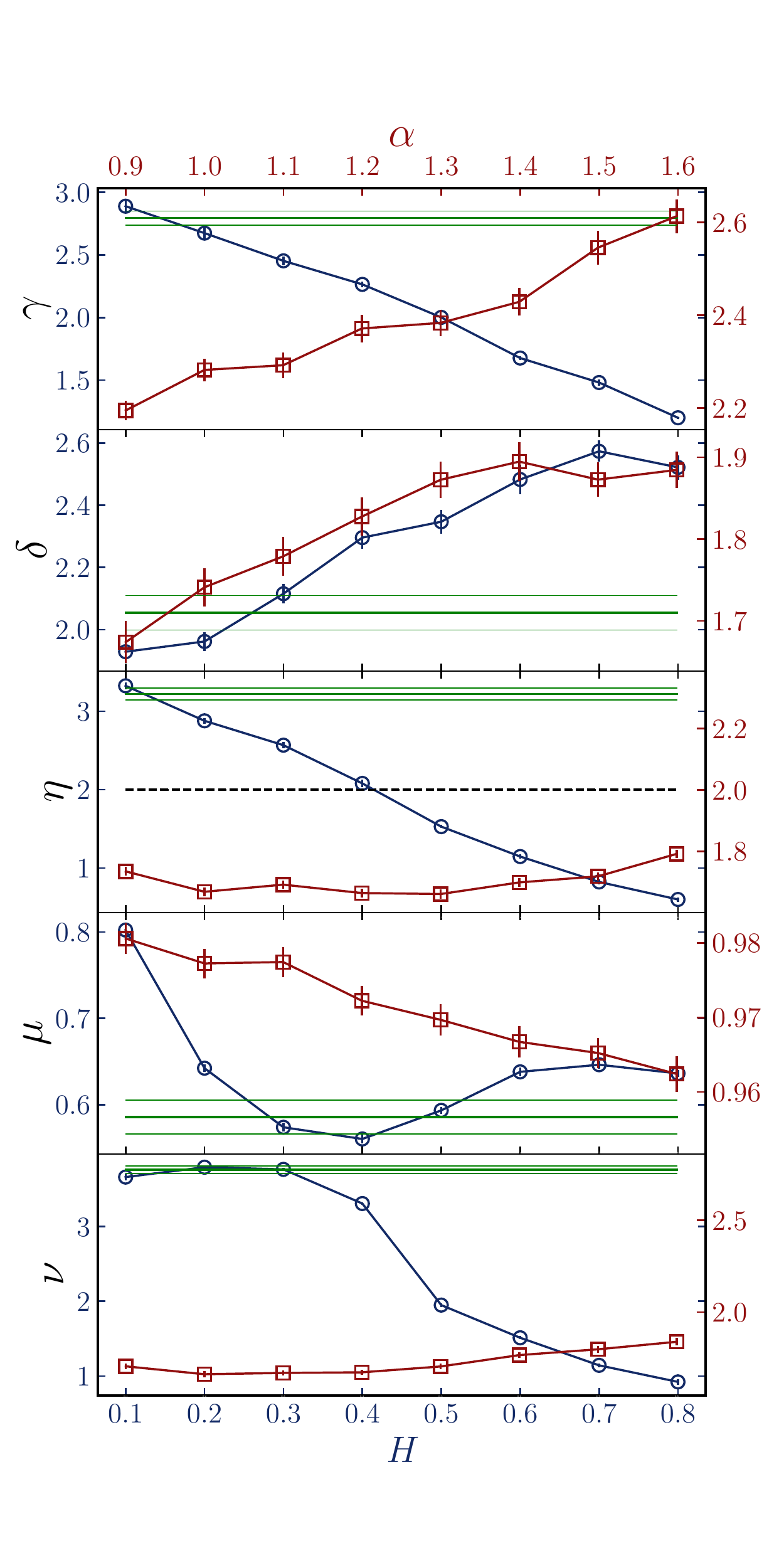}
		\caption{The exponents $\gamma$, $\delta$, $\eta$, $\mu$ and $\nu$ for the SF networks corresponding to three process: the (green) lines show the BTW model and the error bars, the (blue) circles are for FBM, and the (red) square symbols show the Levy results. The left vertical axis shows the exponent values for the FBM series, while the right hand side vertical axis stands for the BTW model and the Levy walk. In the upper (lower) panel the horizontal axis $\alpha$ ($H$) is shown for the Levy walk (FBM series).}
		\label{fig:Exponents}
	\end{figure}
	As a more systematic inspection, we calculate these exponents for FBM and Levy processes, the results of which are shown in Fig.~\ref{fig:Exponents} in the thermodynamics limit. We see that the exponents run with $H$ and $\alpha$, respectively. $\gamma$ is a decreasing (an increasing) function of $H$ ($\alpha$) for FBM (Levy process) VGs. Our analysis shows that the best fitting to the numerical data in the limit $N\rightarrow\infty$ is $\gamma(H)=(3.17\pm 0.03)-(2.42\pm 0.06) H$ for the FBM, which is served as an enhancement of the previously observed relation~\cite{lacasa2009visibility, ni2009degree}. The linear fitting of $\gamma$ in terms of $\alpha$ reveals furthermore that $\gamma(\alpha)=(1.69\pm 0.04)+(0.55\pm 0.05) \alpha$ for the Levy process ($0.9\le \alpha\le 1.6$). The monotonic increase of $\gamma$ in terms of $\alpha$ is understood given the fact that $\alpha$ controls the rare events in the Levy process, and rare events influence the visibility pattern of the nodes in VG. More precisely, $\alpha$ diminishes the abundance of rare events, which itself enhances the visibility conditions of the nodes, so that the degree of nodes with small $k$ values increase, while it decreases for the nodes with large degrees (hubs), giving rise to an increase in $\gamma$, which is shown to be linear. Generally, one expects that the betweenness increases by decreasing $H$ since for small $H$ values the VGs are more sparse. The exponent $\delta$ decreases with decreasing $H$, showing that this increase is smaller for the nodes with smaller betweenness than that for the nodes with larger betweenness. The same argument holds for $\alpha$. \\
	
	In Fig.~\ref{fig:Exponents}c the horizontal dashed-line shows the limit given by the Ref.~\cite{barthelemy2004betweenness}, i.e. $\eta_{\text{max}}=2$ for SFTs. From this figure we see that for the FBM ($\eta_{\text{FBM}}$) in the anticorrelated regime $0<H \lesssim 0.5$ (where the VGs become sparse due to the bad visibility conditions~\cite{masoomy2021persistent}), the conjecture of Eq.~\ref{Eq:Claim} $\text{C}\mathbb{I}$ is violated, i.e. $\eta>\eta_{\text{max}}$ just like the BTW model. For the Levy process, $\eta$ is always smaller than $2$ for all $\alpha$ values in agreement with the conjecture~\cite{barthelemy2004betweenness}. \\
	
	\begin{figure}
		\includegraphics[width=80mm]{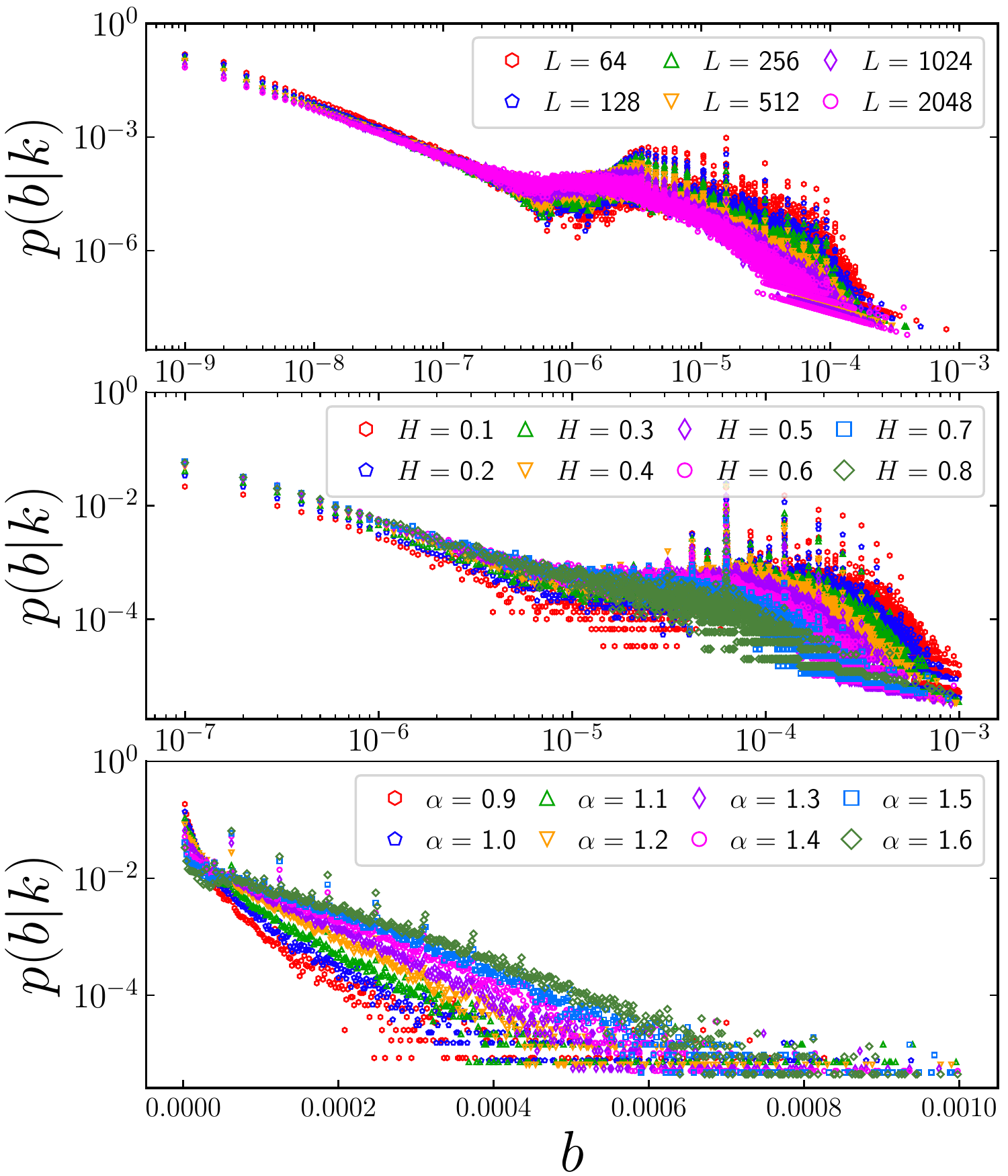}
		\caption{The conditional probability distribution function $p(b|k)$ for various processes: BTW model (top panel), FBM series [$0.1\le H\le 0.8$] (middle panel) and Levy walk [$0.9\le \alpha\le 1.6$] (bottom panel).}
		\label{fig:conditionalP}
	\end{figure}
	
\begin{figure}
		\includegraphics[width=80mm]{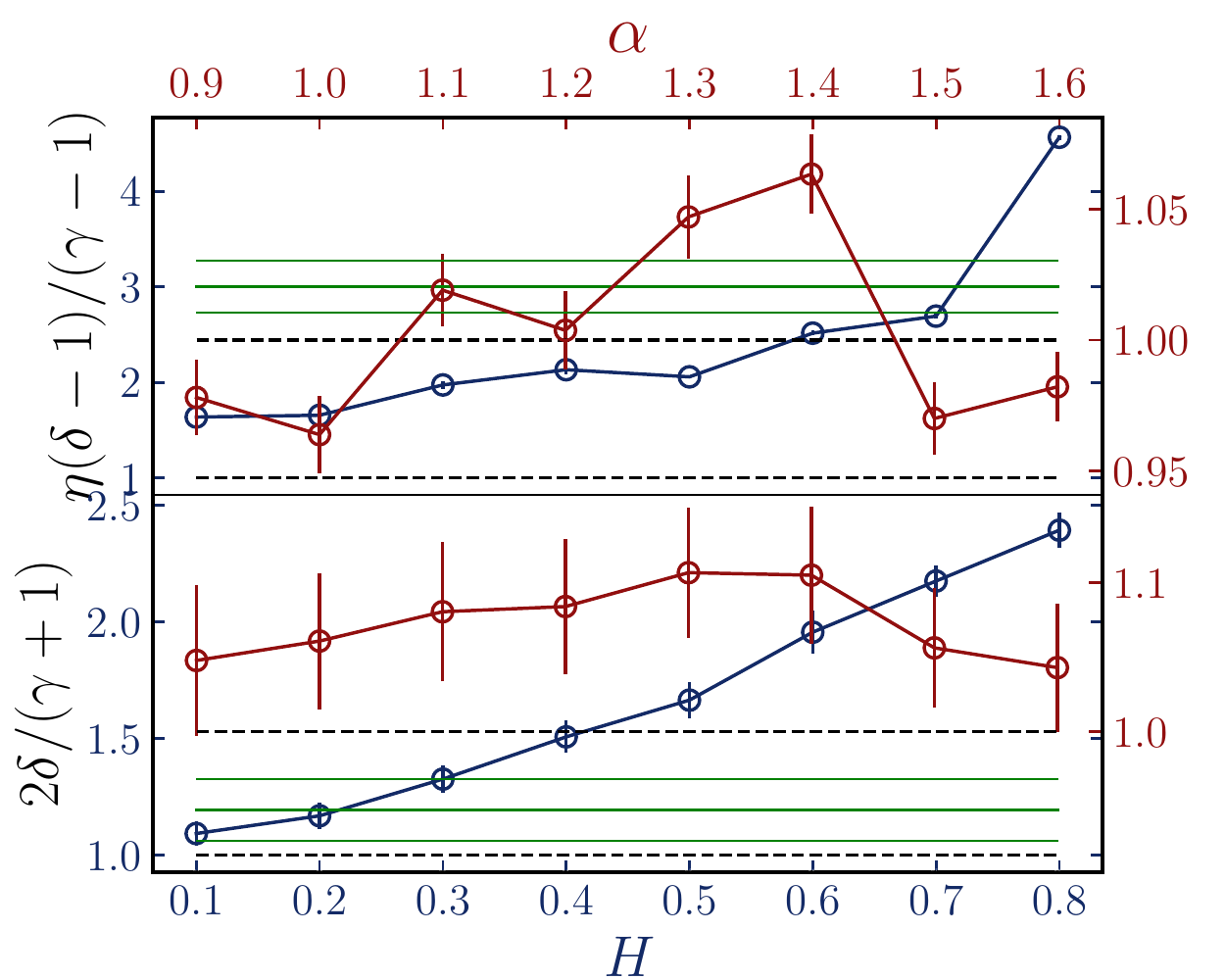}
		\caption{Top panel: hyperscaling relation analysis for various models. The BTW model, FBM series (for all value of $H$) and the Levy walk for some $\alpha$s do not satisfy this relation. Bottom panel: the inequality $\delta\ge\frac{\gamma+1}{2}$ is valid for both FBM series and Levy walk, while it is invalid for BTW model. The left vertical axis is devoted for FBM and the right one is for both BTW and Levy.}
		\label{fig:Hyper}
	\end{figure}

	The reason for this anomalous behavior is the existence of large fluctuations in the scaling $b$-$k$ relation as first pointed out in~\cite{guimera2004modeling, barrat2005effects, kitsak2007betweenness}. This phenomenon leads to some interesting consequences, like the violation of the hyperscaling relation Eq.~\ref{Eq:HyperScaling}, and also the fact that the highest degrees are typically not the most central ones in the sense of betweenness~Ref.~\cite{guimera2004modeling}. It is also responsible for the fractality observed in the synthetic and real-world SF networks~\cite{kitsak2007betweenness}. While, for non-fractal networks, degree and betweenness centralities are strongly correlated, the betweenness centrality of low degree nodes in fractal SF networks can be comparable to that of the hubs~\cite{kitsak2007betweenness}. \\
	Such a large fluctuation should be observed in the conditional probability $p(b|k)$ by inspecting its width. Figure~\ref{fig:conditionalP} shows $p(b|k=10)$ in terms of $b$ for the three cases. Interestingly, we see that this function decays in a power-law (heavy-tail) form for two cases BTW and FBM, while for the Levy process the situation is completely different: it decays exponentially with a finite width avoiding large fluctuations. The exponents for both power-law and exponential decays depend on the correlation parameter ($H$ for FBM and $\alpha$ for Levy). Therefore, one concludes that the width of $p(b|k)$ is finite for the Levy process, the characteristic of the non-fractal SF network, while for the BTW and FBM it is diverging, leading to large fluctuations (a characteristic of fractal SF networks). The violation of the hyperscaling relation Eq.~\ref{Eq:HyperScaling} for the BTW model and FBM (all $H$ values) is shown in the upper graph in Fig.~\ref{fig:Hyper}, while the hyperscaling relation remains almost valid for the Levy process for all $\alpha$ values. For the FBM, while the hyperscaling relation is violated for all $H$ values, the Barthemly's conjecture ($\text{C}\mathbb{I}$, see Fig.~\ref{fig:Hyper}) fails only for $0<H\lesssim 0.5$. Although Fig.~\ref{fig:conditionalP} shows the fluctuations for the smaller values of $H$ are higher (which favors the anomalous behavior), this issue needs some more analysis which is beyond the present paper.\\ 

	Before closing the paper, it is worthy to add notes on the clustering coefficient $c$ as a measure for hierarchical structure in networks, which is a decreasing function of the degree in real-world networks~\cite{vazquez2003topology,ravasz2003hierarchical}. This decrease is power-law for non-tree SF networks $c \propto k^{-\mu}$, where $\mu$ is some exponent, being $1$ for the deactivation model~\cite{klemm2002highly}. This relation is valid for other generalized phenomenological models~\cite{barabasi2001deterministic,dorogovtsev2002pseudofractal,jung2002geometric}, while for the SF networks generated by preferentially attachments $c$ and $k$ are uncorrelated. For the internet network, as a growing SF network, the exponents are time-independent exponents, which are fixed to $\gamma = 2.2 \pm 0.1$ and $\delta = 2.1 \pm 0.2$, and also $\eta \approx 1$ and $\mu \approx 0.75 \pm 0.03$~\cite{vazquez2002large}. Generally, for real-world systems (actor network, language network, the World Wide Web, Internet at the Autonomous System level, which have hierarchical structure) the exponent $\mu$ varies with $\gamma$.
	Many theoretical studied have emerged like the networks based on the Molloy and Reed (MR) algorithm~\cite{molloy2011critical}, generalized BA (GBA)~\cite{albert2000topology} and fitness model~\cite{bianconi2011competition}, with the prediction $\eta_{\rm MR}, \eta_{\rm GBA} \approx 1$ while $\eta_{\rm fitness} \approx 1.4$. For MR and GBA, $c$ does not depend on $k$, while for the fitness model the betweenness decays by the degree in a scaling manner. In Fig.~\ref{fig:Exponents} we show $\mu$ in terms of $\alpha$ and $H$ for Levy and FBM processes, respectively. For the former it is a decreasing function of $\alpha$, while for the FBM it is not monotonic, i.e. the clustering coefficient for hubs decreases leading to larger values for $\mu$, which has not been observed previously. For low $H$ values, the obtained $\mu$ is compatible with the values observed for the Internet network~\cite{vazquez2002large}. \\
	
	To conclude, we considered Barthelemy's conjecture for the betweenness-degree ($b$-$k$) scaling exponent for scale-free (SF) networks, claiming that $\eta_{\text{max}}=2$, belonging to scale-free trees (SFTs), based on which he further conjectured that $\delta\ge\frac{\gamma+1}{2}$. We analyzed the VGs for the time series of the 2D BTW model, 1D FBM (controlled by the Hurst exponent $H$) and 1D Levy walk (controlled by the step-index $\alpha$). We numerically showed that the VGs for all of these models are SF, with well-defined scaling exponents. A super-universal behavior is found for the distribution function for generalized degree function $p(q,t)$ identical to Barabasi-Albert network, see Eq.~\ref{Eq:generalizedQ}. We present pieces of evidence for the violation of Barthelemy's conjecture. Specifically for the BTW model and FBM with $H\lesssim 0.5$, $\eta$ is larger than $2$, and also for the BTW model $\delta<\frac{\gamma+1}{2}$, while Barthelemy's conjecture remains valid for the Levy process for all $\alpha$ values. By analyzing the conditional probability $p(b|k)$ we numerically show that the failure of Barthelemy's conjecture is due to the large fluctuations (or uncertainty) in the $b$-$k$ scaling relation. This function decays in a power-law fashion for the BTW model as well as the FBM for all $H$ values. This results further in a violation of hyperscaling relation $\eta=\frac{\gamma-1}{\delta-1}$ and as a result to some emergent anomalous behaviors as predicted in the literature~\cite{guimera2004modeling, barrat2005effects, kitsak2007betweenness} for the BTW model and FBM series.

	-----------------------
	
	\bibliography{refs}
	
	

	\newpage
	
	\setcounter{equation}{0}
	\setcounter{figure}{0}
	\setcounter{table}{0}
	\renewcommand\theequation{SM\arabic{equation}}
	\renewcommand\thefigure{SM\arabic{figure}}
	\renewcommand\thetable{SM\arabic{table}}
	
	\begin{center}
		\large{\textbf{Supplemental Material}}
	\end{center}

\begin{figure*}
			\includegraphics[width=55mm]{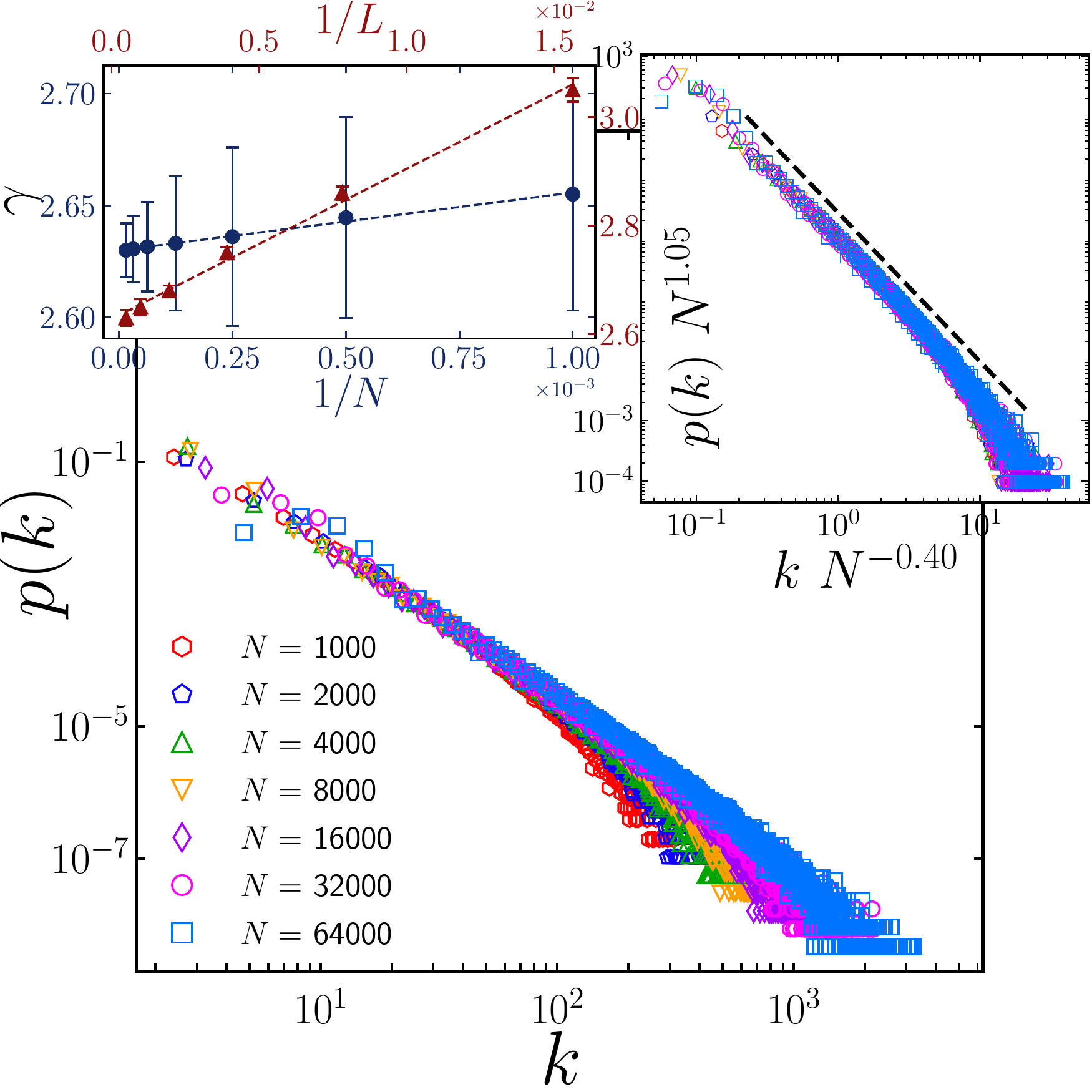}
			\includegraphics[width=55mm]{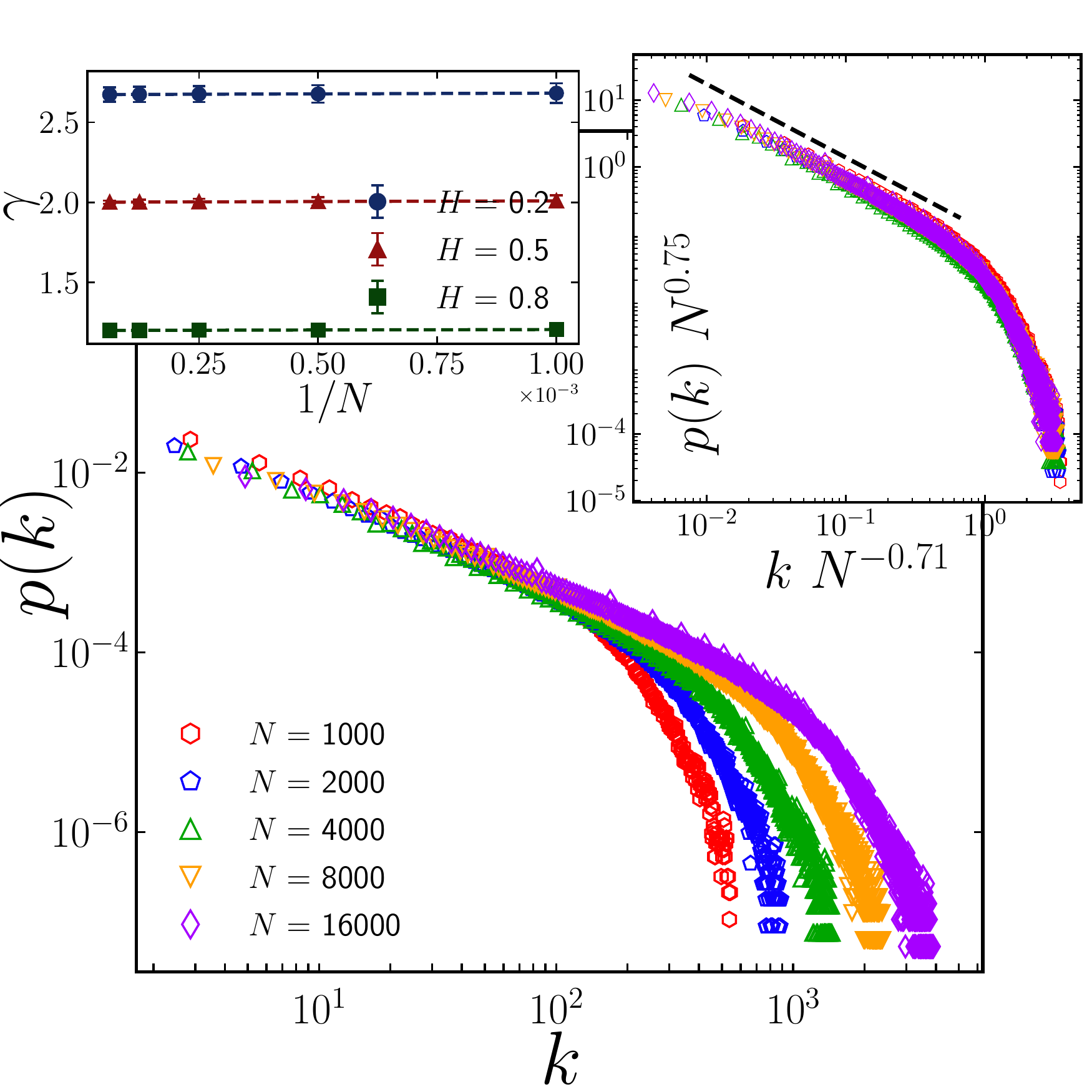}
			\includegraphics[width=55mm]{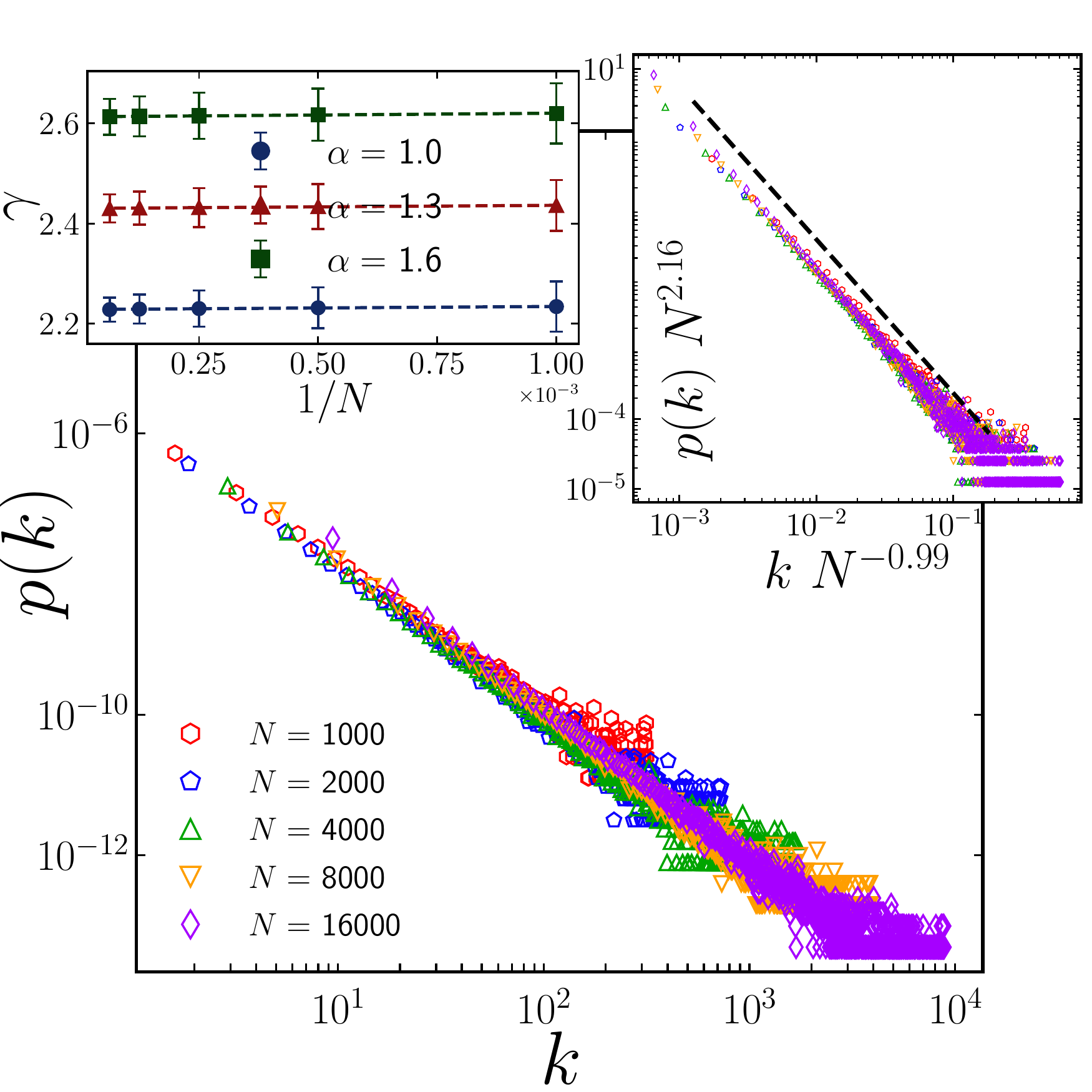}
			\includegraphics[width=55mm]{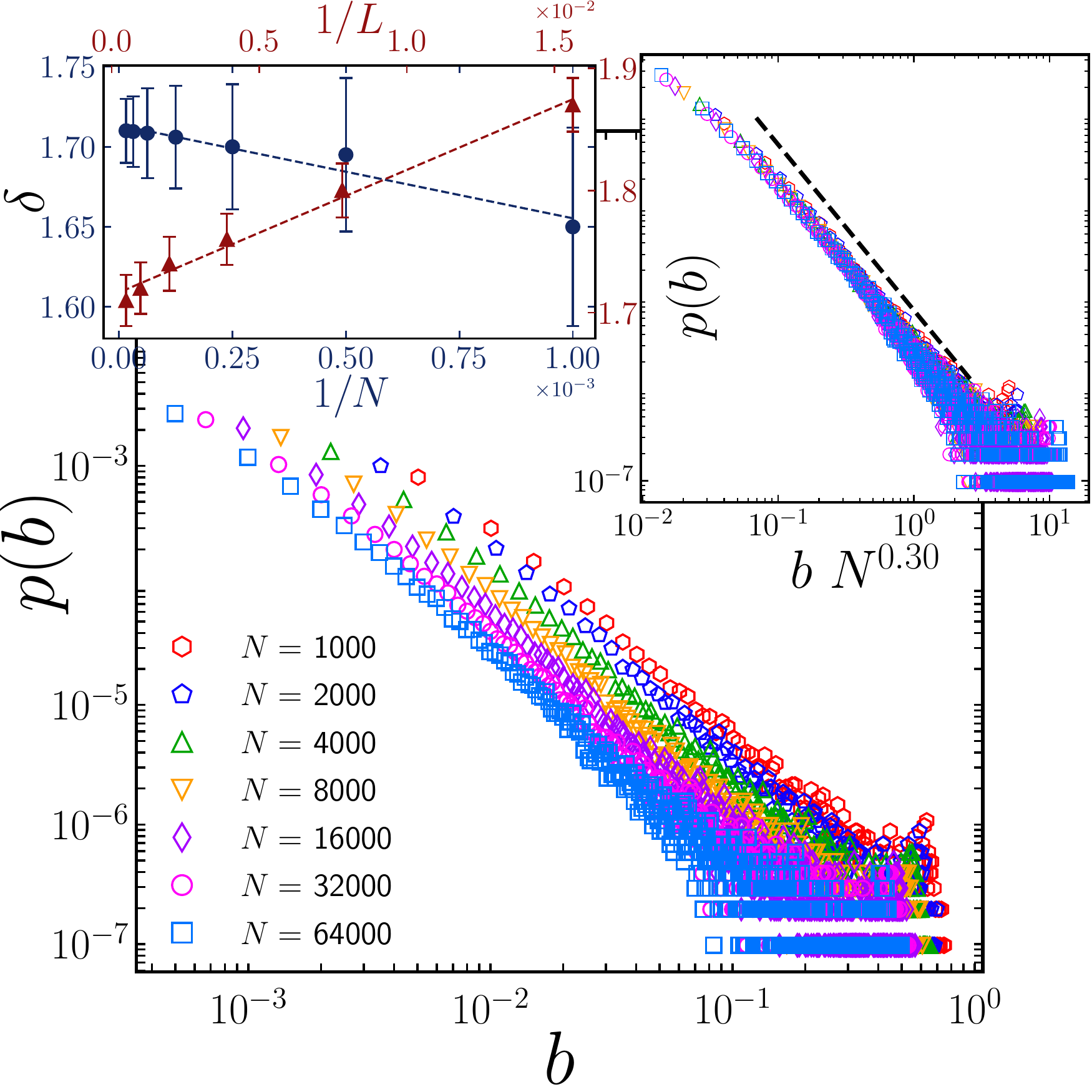}
			\includegraphics[width=55mm]{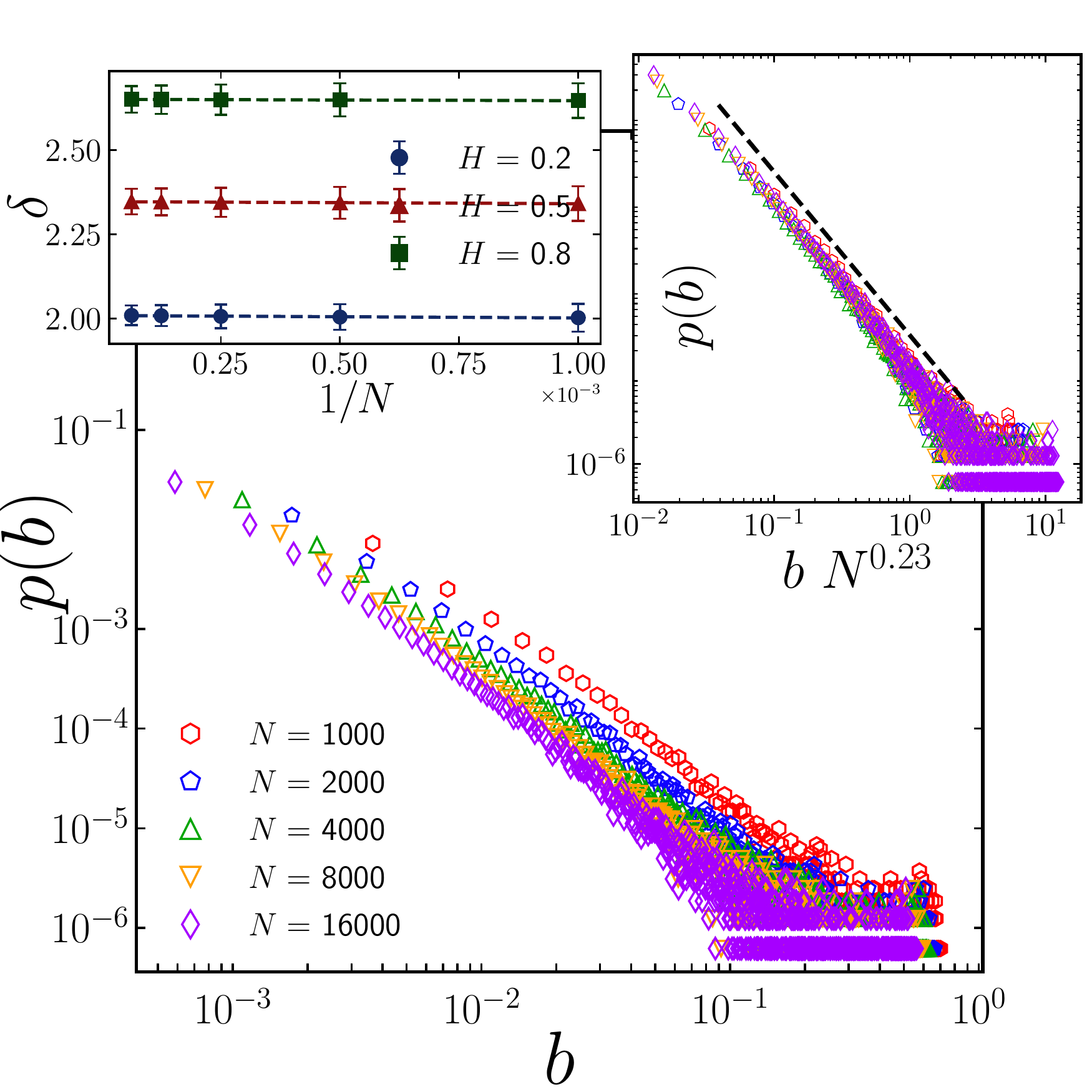}
			\includegraphics[width=55mm]{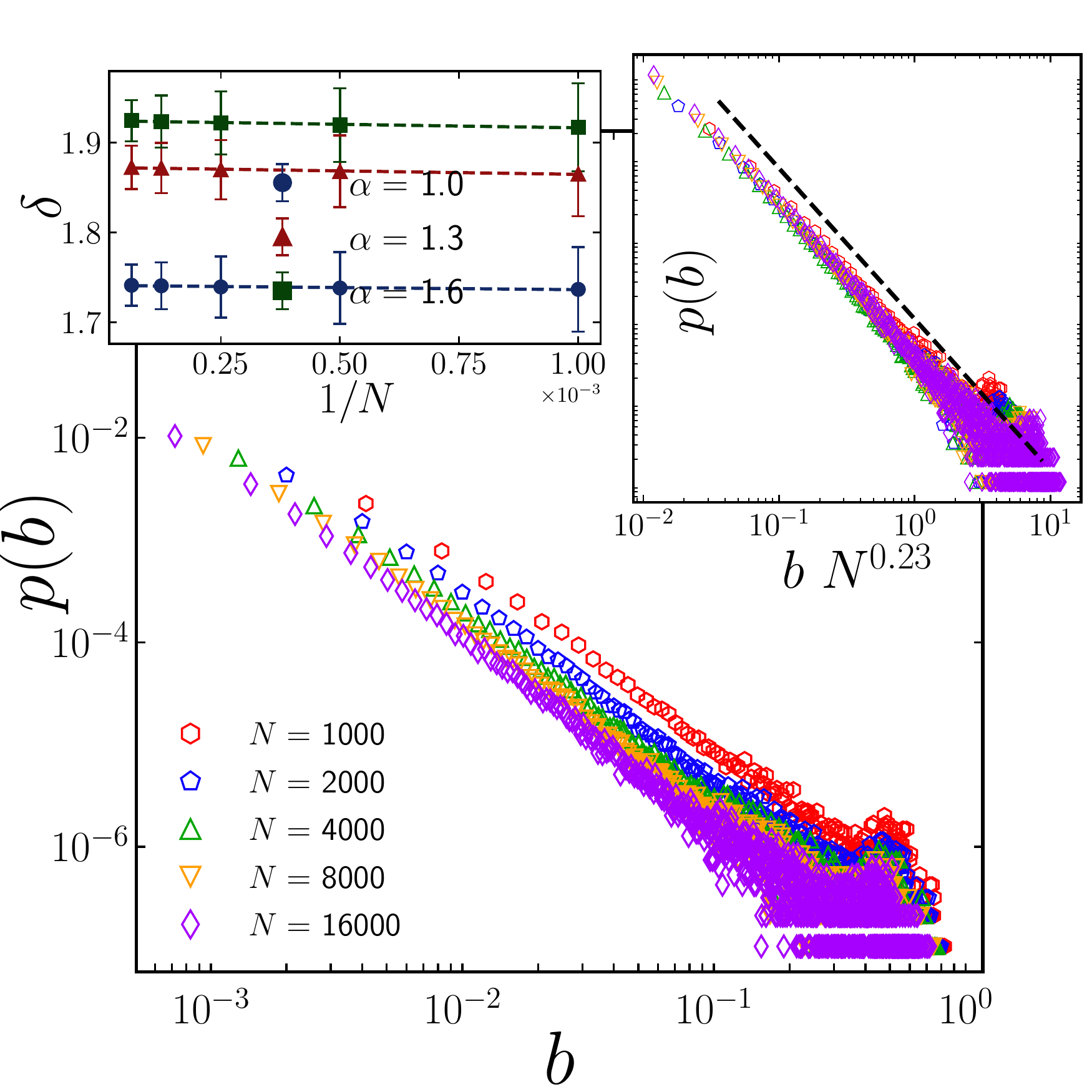}
			\caption{In this figure we show PDF of degree (top row) and betweenness centrality (bottom row) of VG constructed from time series of BTW sandpile model (left column), FBM series (middle column) and Levy walk (right column) for different system size. The scaling behavior of these quantities reveals that the VGs are SF. In the inset plots we show size dependency of the scaling exponent and data collapse analysis as well.}
			\label{fig:Graphs2}
		\end{figure*}

		In this supplementary material, we present various graphs from which the results in Fig.~\ref{fig:Exponents} of the paper were obtained. In the Fig.~\ref{fig:Graphs2} we show various probability density functions (PDFs). The first row shows the PDF of degree for the BTW model (left), FBM series (middle), and Levy processes (right). In the second row, the PDF for betweenness centrality is shown for these models (with the same arrangement as the first row). \\
		
\begin{figure*}
			\includegraphics[width=55mm]{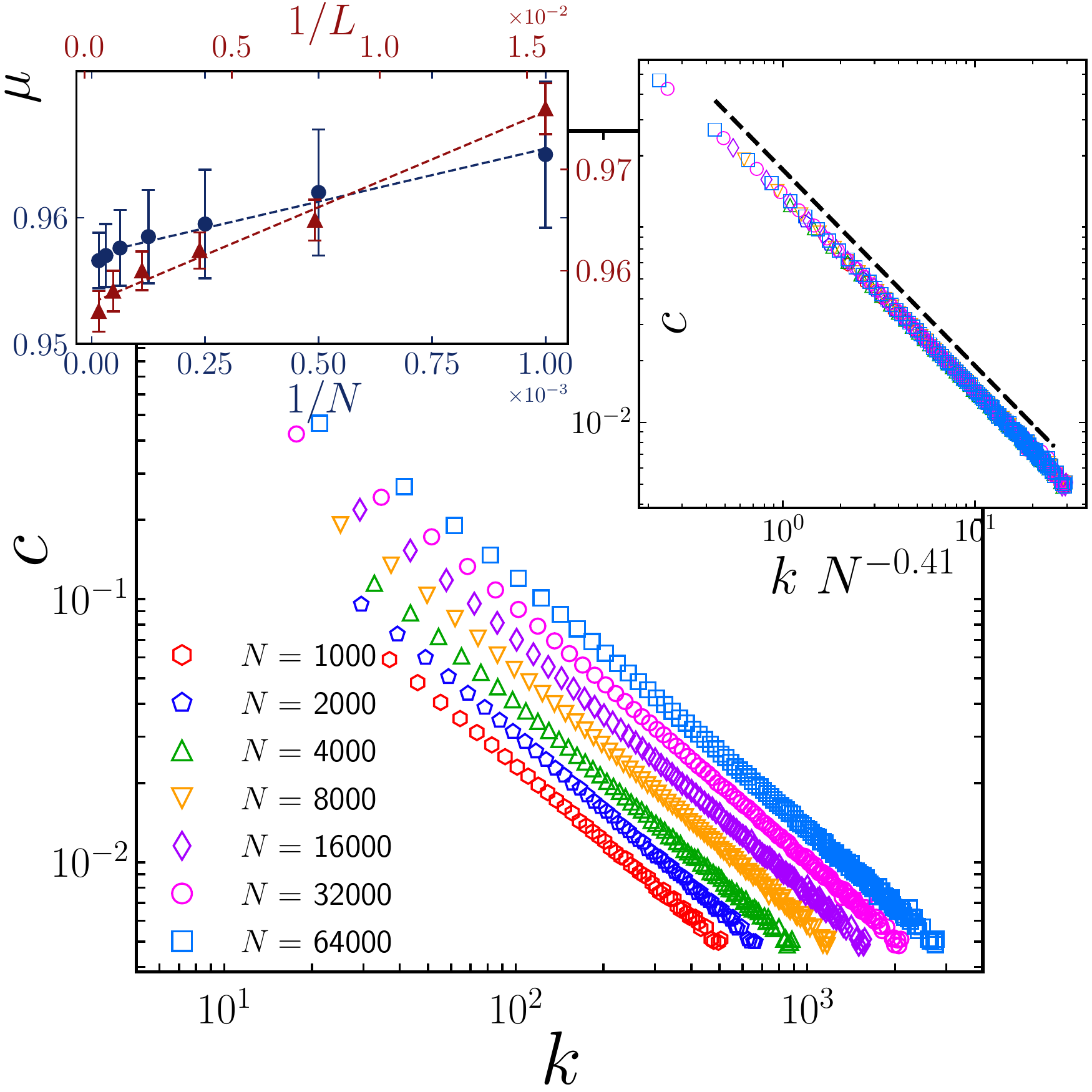}
			\includegraphics[width=55mm]{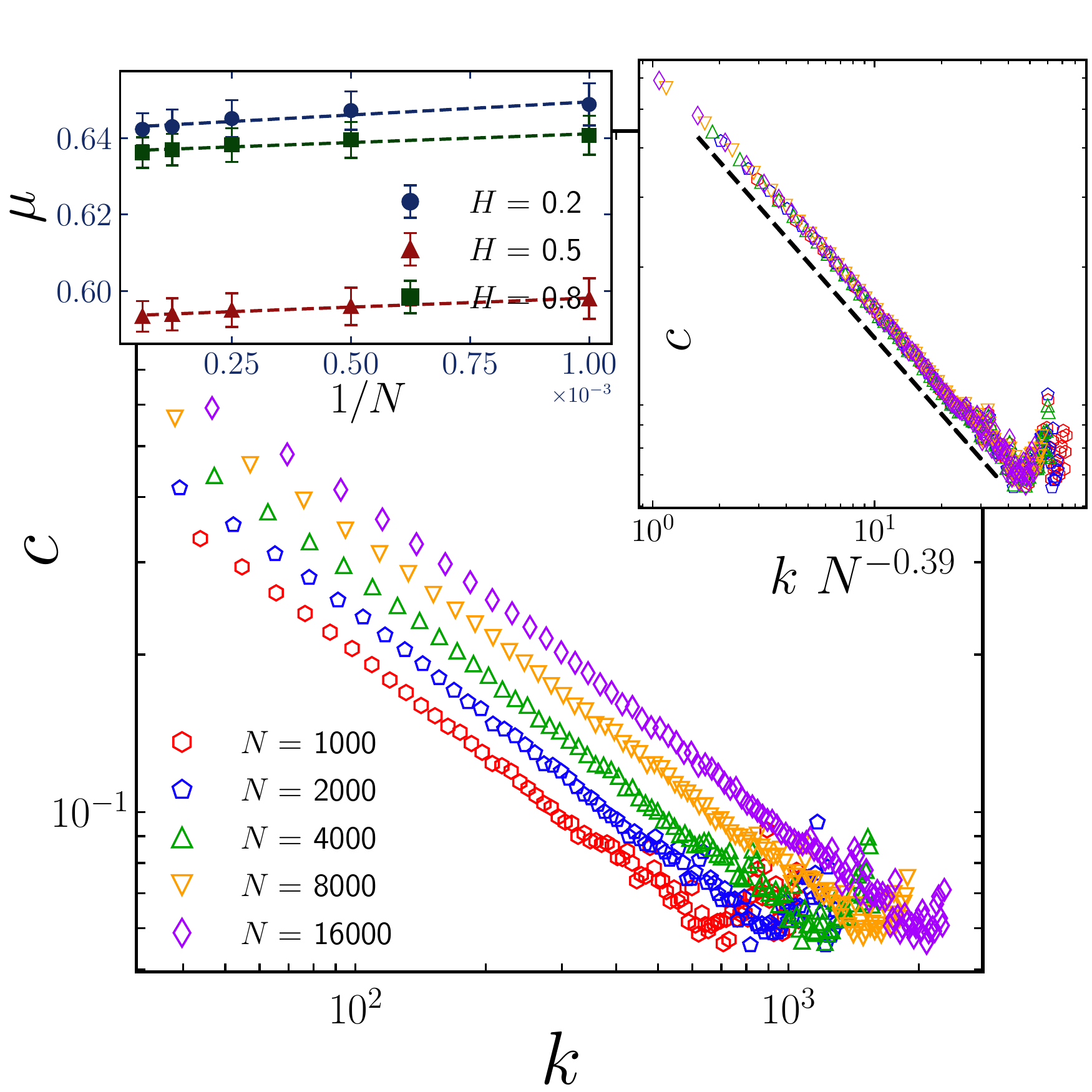}
			\includegraphics[width=55mm]{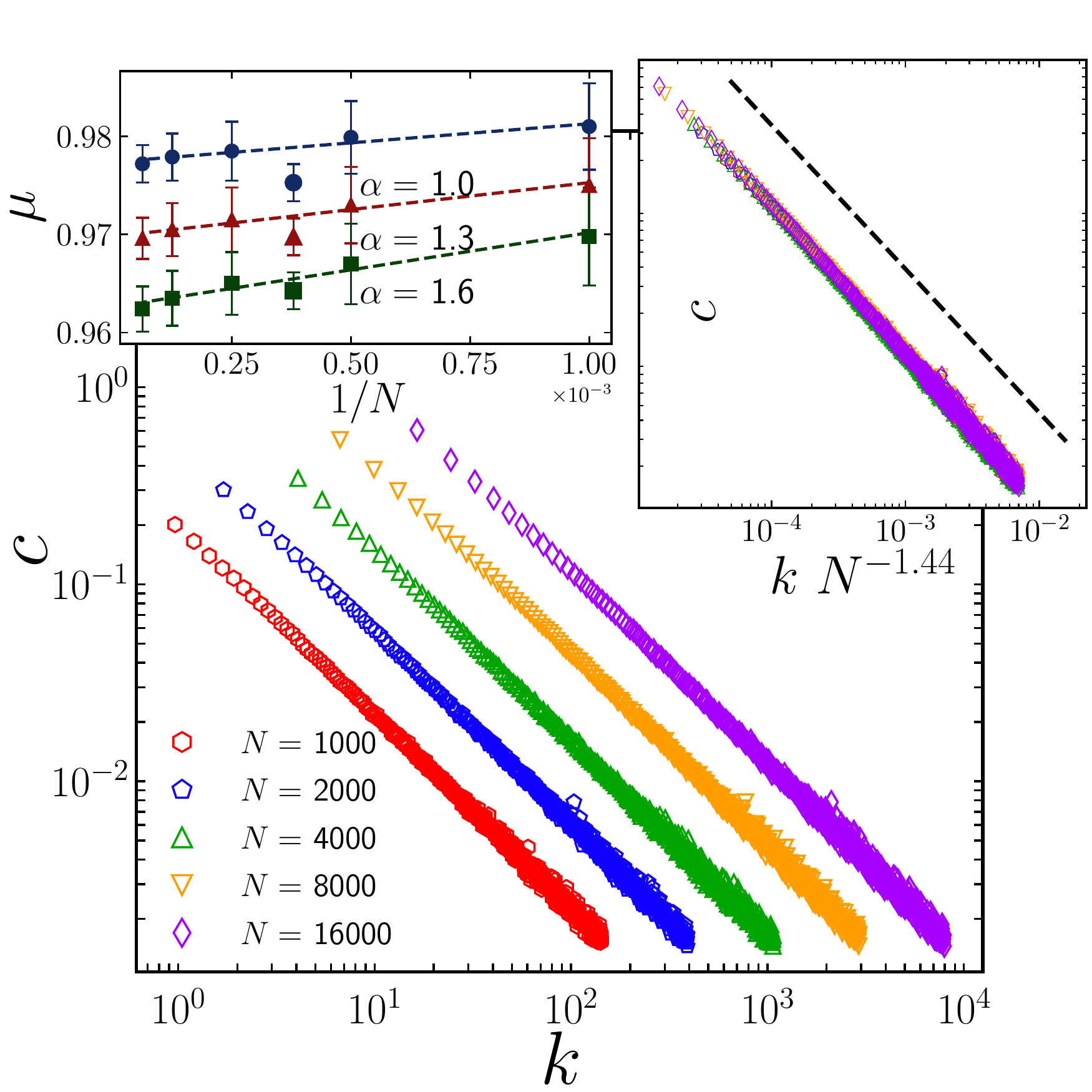}
			\includegraphics[width=55mm]{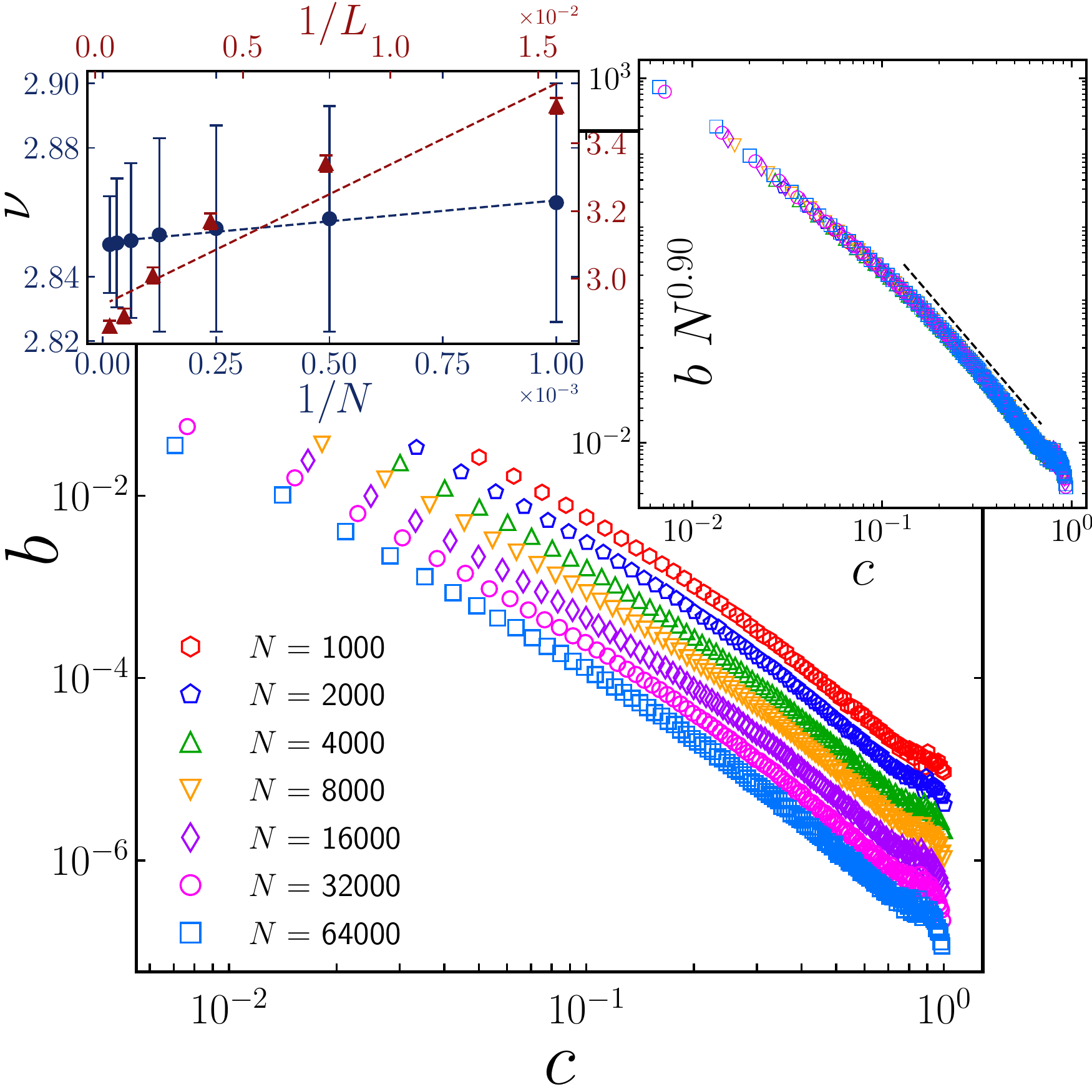}
			\includegraphics[width=55mm]{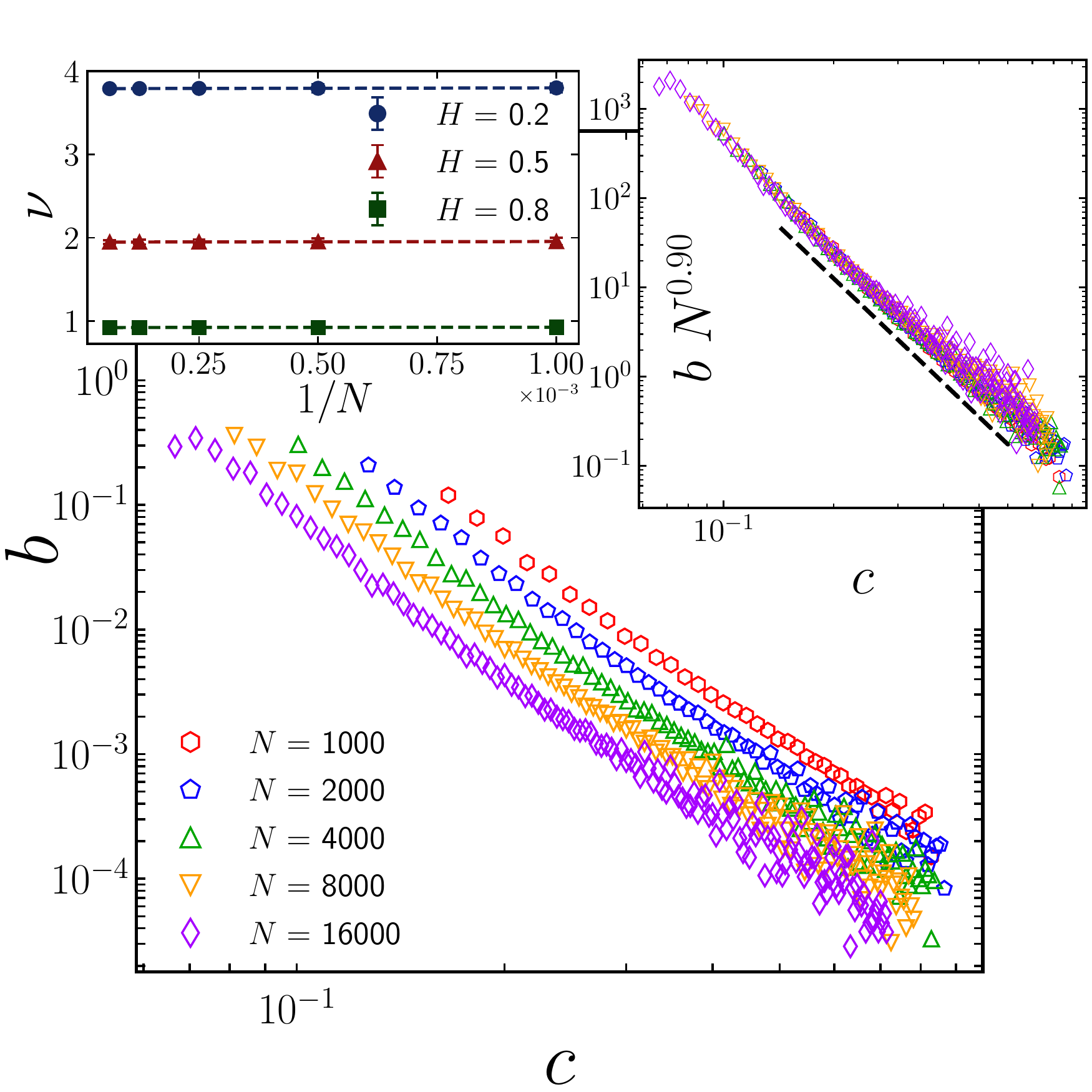}
			\includegraphics[width=55mm]{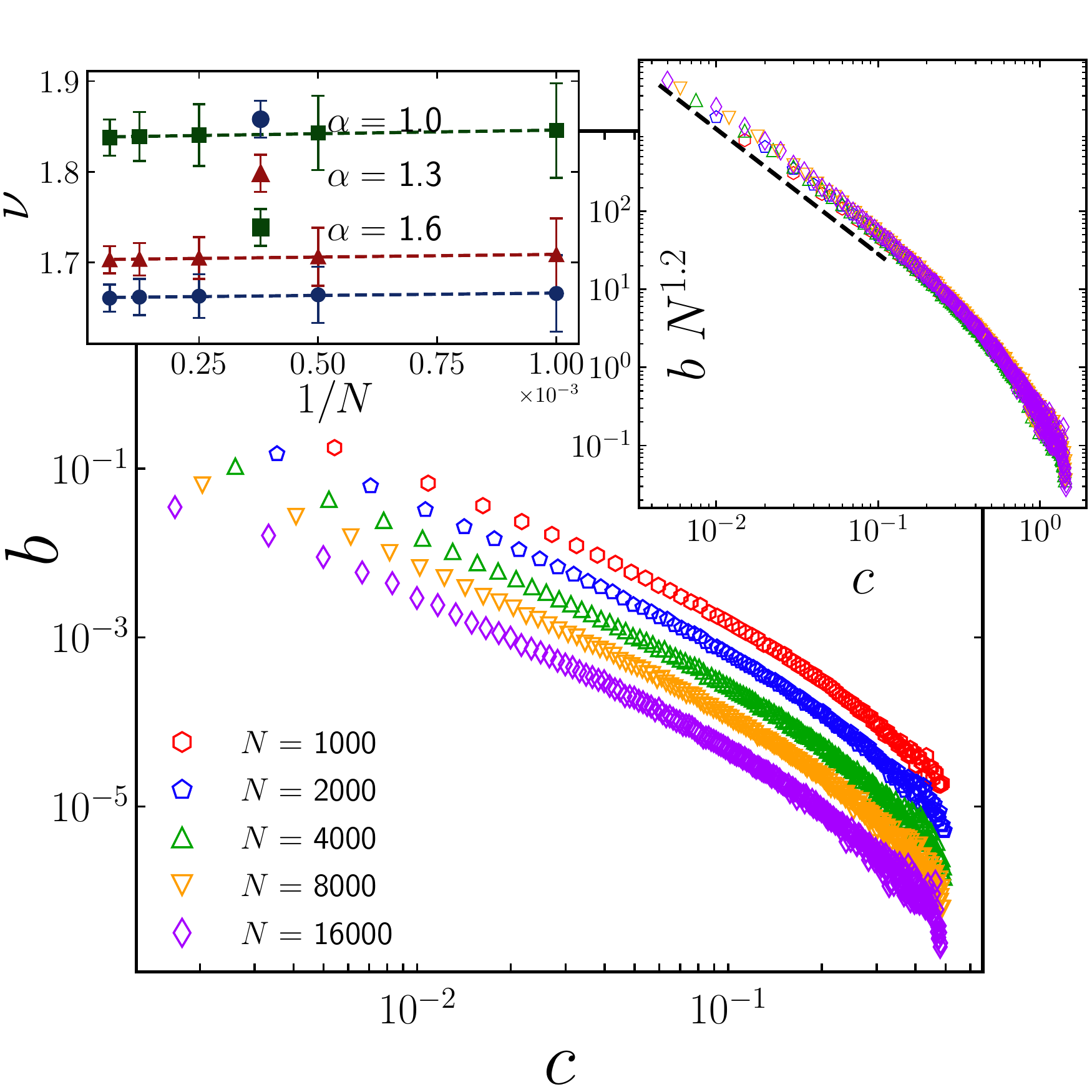}
			\caption{Correlation analysis of clustering coefficient versus degree (top row) and betweenness centrality versus clustering coefficient (bottom row) of VG constructed from time series of BTW sandpile model (left column), FBM series (middle column) and Levy walk (right column) for different system size in log-log scale. In the inset plots we show size dependency of the corresponding exponent and data collapse analysis.}
			\label{fig:Graphs3}
		\end{figure*}

		Fig.~\ref{fig:Graphs3} indicates some other scaling relation between statistical obseravbles, i.e. clustering coefficient ($c$) versus degree ($k$) (the first row) and betweenness centrality ($b$) versus clustering coefficient (the second row) for the BTW model (left), FBM series (middle) and Levy processes (right).

		\begin{table*}
			\begin{tabular}{|c|c|c|c|c|c|c|c|c|c|}
				\hline $\alpha_{k}$ & $\beta_{k}$ & $\alpha_{b}$ & $\beta_{b}$ & $\alpha_{b,k}$ & $\beta_{b,k}$ & $\alpha_{c,k}$ & $\beta_{c,k}$ & $\alpha_{b,c}$ & $\beta_{b,c}$ \\
				\hline $1.05 \pm 0.03$ & $0.40\pm 0.02$ & $-$ & $0.30 \pm 0.03$ & $-$ & $0.50 \pm 0.03$ & $-$ & $0.41\pm 0.02$ & $0.90 \pm 0.02$ & $-$ \\
				\hline
			\end{tabular}
			\caption{In this table we show the re-scaling exponents of various quantities for the BTW model.}
			\label{TAB:exponentsk=2}
		\end{table*}

\end{document}